\documentclass[sigconf]{acmart}

\def\BibTeX{{\rm B\kern-.05em{\sc i\kern-.025em b}\kern-.08emT\kern-.1667em\lower.7ex\hbox{E}\kern-.125emX}}
    
\acmYear{2020}\copyrightyear{2020}
\setcopyright{acmlicensed}
\acmConference[FAT* '20]{Conference on Fairness, Accountability, and Transparency}{January 27--30, 2020}{Barcelona, Spain}
\acmBooktitle{Conference on Fairness, Accountability, and Transparency (FAT* '20), January 27--30, 2020, Barcelona, Spain}
\acmPrice{15.00}
\acmDOI{10.1145/3351095.3372862}
\acmISBN{978-1-4503-6936-7/20/01}

\usepackage{subcaption} 
\usepackage{wrapfig}
\usepackage{booktabs}
\usepackage{tabularx}
\usepackage{lipsum}
\usepackage{adjustbox}

\newcommand{\code}[1]{\texttt{#1}}
\begin{document}

\title[Garbage In, Garbage Out?]{Garbage In, Garbage Out? Do Machine Learning Application Papers in Social Computing Report Where Human-Labeled Training Data Comes From?}


\settopmatter{authorsperrow=4}
\author{R. Stuart Geiger}
\orcid{0000-0001-7215-0532}
\authornote{Corresponding author: stuart@stuartgeiger.com}
\affiliation{ \institution{University of California, Berkeley}}
\author{Kevin Yu}
\affiliation{\institution{University of California, Berkeley}}

\author{Yanlai Yang}
\affiliation{ \institution{University of California, Berkeley}}

\author{Mindy Dai}
\affiliation{ \institution{University of California, Berkeley}}

\author{Jie Qiu}
\affiliation{ \institution{University of California, Berkeley}}

\author{Rebekah Tang}
\affiliation{ \institution{University of California, Berkeley}}

\author{Jenny Huang}
\affiliation{ \institution{University of California, Berkeley}}

\renewcommand{\shortauthors}{Geiger et al.}

%
\begin{abstract}
Many machine learning projects for new application areas involve teams of humans who label data for a particular purpose, from hiring crowdworkers to the paper's authors labeling the data themselves. Such a task is quite similar to (or a form of) structured content analysis, which is a longstanding methodology in the social sciences and humanities, with many established best practices. In this paper, we investigate to what extent a sample of machine learning application papers in social computing --- specifically papers from ArXiv and traditional publications performing an ML classification task on Twitter data --- give specific details about whether such best practices were followed. Our team conducted multiple rounds of structured content analysis of each paper, making determinations such as: Does the paper report who the labelers were, what their qualifications were, whether they independently labeled the same items, whether inter-rater reliability metrics were disclosed, what level of training and/or instructions were given to labelers, whether compensation for crowdworkers is disclosed, and if the training data is publicly available.  We find a wide divergence in whether such practices were followed and documented. Much of machine learning research and education focuses on what is done once a ``gold standard'' of training data is available, but we discuss issues around the equally-important aspect of whether such data is reliable in the first place.
\end{abstract}

%
%



\maketitle


\section{Introduction}

Machine learning (ML) has become widely used in many academic fields, as well as across the private and public sector. Supervised machine learning is particularly prevalent, in which training data is collected for a set of entities with known properties (a ``ground truth'' or ``gold standard''), which is used to create a classifier that will make predictions about new entities of the same type. Supervised ML requires high-quality training data to produce high-quality classifiers. ``Garbage In, Garbage Out'' is a longstanding aphorism in computing about how flawed input data or instructions will produce flawed outputs. \cite{mellin1957work,babbage2011passages} However, contemporary ML research and education tends to focus less on obtaining and validating such a training dataset, with such considerations often passed over in major textbooks \cite[e.g.][]{friedman2009elements,james2013introduction,goodfellow2016deep}.   The predominant focus is typically on what is done with the training data to produce a classifier, with heavy emphasis on mathematical foundations and routine use of clean and tidy ``toy'' datasets. The process of creating a ``gold standard'' or ``ground truth'' dataset is routinely black-boxed. Many papers in ML venues are expected to use a standard, public training dataset, with authors comparing various performance metrics on the same dataset. While such a focus on what is done to a training dataset may be appropriate for theoretically-oriented basic research in ML, this is not the case for supervised ML applications.

\subsection{Study overview}

All approaches of producing a training dataset involve some form of human judgment, albeit at varying levels of granularity. In this paper, we investigate and discuss a wide range of issues and concerns around the curation of human-labeled or human-annotated data, in which one or more individuals make discrete assessments of items. We report from a study in which a team of six labelers systematically examined a corpus of supervised machine learning application papers in social computing, specifically those that classified tweets from Twitter for various purposes. For each paper, we recorded what the paper does or does not state about the training data used to produce the classifier presented in the paper. The bulk of the papers we examined were a sample of preprints or postprints published on ArXiV.org, plus a smaller set of published papers sampled from Scopus. We determined whether such papers involved an original classification task using supervised ML, whether the training data labels were produced from human annotation, and if so, the source of the human-labeled dataset (e.g. the paper's authors, Mechanical Turk, recruited experts, no information given, etc.). For all papers in which an original human-labeled dataset was produced, we then made a series of further determinations, including if definitions and/or examples were given to labelers, if labelers independently labeled the same items, if inter-rater reliability metrics were presented, if compensation details for crowdworkers were reported, if a public link to the dataset was available, and more. 

As our research project was a human-labeling project studying other human-labeling projects, we took care in our own practices. We only have access to the paper reporting about the study and not the actual study itself, and many papers either do not discuss such details at all or without sufficient detail to make a determinations. For example, many papers did note that the study involved the creation of an original human-labeled dataset, but did not specify who labeled it. For some of our items, one of the most common labels we gave was ``no information'' --- which is a concerning issue, given how crucial such information is in understanding the validity of the training dataset and by extension, the validity of the classifier.

\section{Literature review and motivation}

\subsection{A different kind of ``black-boxing'' in machine learning}

In the introduction, we noted training data is frequently black-boxed in machine learning research and applications. We use the term ``black-boxed'' in a different way than it is typically invoked in and beyond the FAT* community, where often refers to interpretability. In that sense, ``black-boxing'' means that even for experts who have access to the training data and code which created the classifier, it is difficult to understand why the classifier made each decision. In social science and humanities work on ``black-boxing'' of ML (and other ``algorithmic'' systems), there is often much elision between issues of interpretability and intentional concealment, as Burrell \cite{burrell2016machine} notes. A major focus is on public accountability \cite[e.g.][]{pasquale2015black}, where many problematic issues can occur behind closed doors. This is even the case with relatively simple forms of analytics and automation --- such as if-then statements, linear regressions, or rule-based expert systems \cite{stuart2004databases,eubanks2018automating}. 

In contrast, we are concerned with what is and is not taken for granted when developing a classifier. This use is closer to how Latour \& Woolgar used it in an ethnographic study of scientific laboratories \cite{latour1979laboratory}. They discuss how equipment like a mass spectrometer would typically be implicitly trusted to turn samples into signals. However, when the results were drastically unexpected, it could be a problem with the machine or a fundamental breakthrough. Scientists and technicians would have to ``open up the black box,'' changing their relationship to the equipment to determine if the problem was with the equipment or the prevailing theory. In this view, black-boxing is a relational concept, not an objective property. It is about the orientation people have to the same social-technical systems they routinely work with and rely upon. ``Opening up the black box'' is not about digging into technical or internal details \textit{per se}, but a gestalt shift in whether the output of a system is implicitly taken for granted or open for further investigation. 

In this view, black-boxing is not inherently problematic. The question is more about who gets to be skeptical about data and who is obligated to suspend disbelief, which are also raised in discussions of open science \& reproducibility \cite{Kitzes2018}. Operationalization, measurement, and construct validity have long been crucial and contested topics in the social sciences. Within quantitative sub-fields, it is common to have extensive debates about the best way to define and measure a complex concept (e.g. ``intelligence''). From a qualitative and Science \& Technology Studies perspective, there is extensive work on the practices and implications of various regimes of measurement \cite{Goodwin1994,scott_seeing_1998,Latour1999a,bowker1999sorting}. In ML, major operationalization decisions can implicitly occur in data labeling. Yet as Jacobs \& Wallach note, ``[i]n computer science, it is particularly rare to articulate the distinctions between constructs and their operationalizations'' \cite[p. 19]{jacobs_measurement_2019}. This is concerning, because ``many well-studied harms [in ML] are direct results of a mismatch between the constructs purported to be measured and their operationalizations'' \cite[p. 14]{jacobs_measurement_2019}. 

\subsection{Content analysis}
Creating human-labeled training datasets for machine learning often looks like \textit{content analysis}, a well-established methodology in the humanities and the social sciences (particularly literature, communication studies, and linguistics), which also has versions used in the life, ecological, and medical sciences. Content analysis has taken many forms over the past century, from more positivist methods that formally establish structural ways of evaluating content to more interpretivist methods that embrace ambiguity and multiple interpretations, such as grounded theory \cite{glaser}. The intersection of ML and interpretivist approaches is outside of the scope of this article, but it is an emerging area of interest \cite[e.g.][]{nelson}.

Today, structured content analysis (also called ``closed coding'') is used to turn qualitative or unstructured data of all kinds into structured and/or quantitative data, including media texts, free-form survey responses, interview transcripts, and video recordings. Projects usually involve teams of ``coders'' (also called ``annotators'', ``labelers'', or ``reviewers''), with human labor required to ``code'', ``annotate'', or ``label'' a corpus of items. (Note that we use such terms interchangeably in this paper.) In one textbook, content analysis is described as a ``systematic and replicable'' \cite[p. 19]{riff2013analyzing} method with several best practices: A ``coding scheme'' is defined, which is a set of labels, annotations, or codes that items in the corpus may have. Schemes include formal definitions or procedures, and often include examples, particularly for borderline cases. Next, coders are trained with the coding scheme, which typically involves interactive feedback. Training sometimes results in changes to the coding scheme, in which the first round becomes a pilot test. Then, annotators independently review at least a portion of the same items throughout the entire process, with a calculation of ``inter-annotator agreement'' or ``inter-rater reliability.'' Finally, there is a process of ``reconciliation'' for disagreements, which is sometimes by majority vote without discussion and other times discussion-based.

Structured content analysis is a difficult, complicated, and labor-intensive process, requiring many different forms of expertise on the part of both the coders and those who manage them. Historically, teams of students have often performed such work. With the rise of crowdwork platforms like Amazon Mechanical Turk, crowdworkers are often used for content analysis tasks, which are often similar to other kinds of common crowdworking tasks. Google's reCAPTCHA \cite{von2008recaptcha} is a Turing test in which users perform annotation tasks to prove their humanness --- which initially involved transcribing scanned phrases from books, but now involves image labeling for autonomous vehicles. There are major qualitative data analysis software tools that scaffold the content analysis process to varying degrees, such as MAXQDA or NVivo, which have support for inter-annotator agreement metrics. There have also been many new software platforms developed to support more micro-level annotation or labeling at scale, including in citizen science, linguistics, content moderation, and more general-purpose use cases \cite{chang_revolt_2017,maeda_annotation_2008,perez_marky_2015,bontcheva_gate_2013,halfaker2019ores,doccano}. For example, the Zooniverse \cite{Simpson2014} provides a common platform for citizen science projects across different domain application areas, which let volunteers make judgements about items, which are aggregated and reconciled in various ways.

\subsection{Meta-research and methods papers in linguistics and crowdsourcing}
\label{sec:mrmethods}
Our paper is also in conversation with various meta-research and standardization efforts in linguistics, crowdsourcing, and other related disciplines. Linguistics and Natural Language Processing have long struggled with issues around standardization and reliability of linguistic tagging. Linguistics researchers have long developed best practices for corpus annotation \cite[e.g.][]{hovy2010towards}, including recent work about using crowdworkers \cite{sabouetal2014}. Annotated corpus projects often release guidelines and reflections about their process. For example, the Linguistic Data Consortium's guidelines for annotation of English-language entities (version 6.6) is 72 single-spaced pages \cite{linguistic2008ace}. A universal problem of standardization is that there are often too many standards and not enough enforcement. As \cite{bender2018data} notes, 33-81\% of linguistics/NLP papers in various venues do not even mention the name of the language being studied (usually English). A meta-research study found only 1 in 9 qualitative papers in Human-Computer Interaction reported inter-rater reliability metrics \cite{McDonald2019}.


Another related area are meta-research and methods papers focused on identifying or preventing low-effort responses from crowdworkers --- sometimes called ``spam'' or ``random'' responses, or alternatively ''fraudsters'' or ''cheaters.'' Rates of ``self-agreement'' are often used, determining if the same person labels the same item differently at a later stage. One paper \cite{Mozetic2016} examined 17 crowdsourced datasets for sentiment analysis and found none had self-agreement rates (Krippendorf's alpha) above 0.8, with some lower than 0.5. Another paper recommends the self-agreement strategy in conjunction with asking crowdworkers to give a short explanation of their response, even if the response is never actually examined. \cite{soberon2013measuring}. One highly-cited paper \cite{raykar2012eliminating} proposes a strategy in which crowdworkers are given some items with known labels (a gold/ground truth), and those who answer incorrectly are successively given more items with known labels, with a Bayesian approach to identifying those who are answering randomly. 

\subsection{The data documentation movements}
\label{sec:datadoc}
Our paper is also in conversation with two related movements in computationally-supported knowledge production that have surfaced issues around documentation. First, we see connections with the broader open science and reproducibility movements. Open science is focused on a range of strategies, including open access research publications, educational materials, software tools, datasets, and analysis code \cite{fecher_open_2014}. The reproducibility movement is deeply linked to the open science movement, focusing on getting researchers to release everything that is necessary for others to perform the same tasks needed to get the exact same results \cite{Wilson2017, Kitzes2018}. This increasingly includes pushing for high standards for releasing protocols, datasets, and analysis code. As more funders and journals are requiring releasing data, the issue of good documentation for data and protocols is rising \cite{Goodman2014,gil_toward_2016}. There are also intersecting literatures on systems for capturing information in ML data flows and supply chains \cite{singh_decision_2019,schelter_automatically_2017,gharibi_automated_2019}, as well as supporting data cleaning \cite{schelter_automating_2018,krishnan_activeclean_2016}. These issues have long been discussed in the fields of library and information science, particularly in Research Data Management \cite{schreier2006academic, borgman2012conundrum, Medeiros2017, sallans2012dmp}. 

A major related movement is in and around the FATML field, with many recent papers proposing training data documentation in the context of ML. Various approaches, analogies, and metaphors have been taken in this area, including ``datasheets for datasets'' \cite{gebru2018datasheets}, ''model cards'' \cite{mitchell2019model}, ``data statements'' \cite{bender2018data}, ``nutrition labels'' \cite{holland2018dataset}, a ``bill of materials'' \cite{barclay2019towards}, ``data labels'' \cite{beretta2018ethical}, and ``supplier declarations of conformity'' \cite{hind2018increasing}. Many go far beyond the concerns we have raised around human-labeled training data, as some are also (or primarily) concerned with documenting other forms of training data, model performance and accuracy, bias, considerations of ethics and potential impacts, and more. We discuss how our findings relate to this broader emerging area more in the concluding discussion.

\section{Data and methods}
\subsection{Data: machine learning papers performing classification tasks on Twitter data}
Our goal was to find a corpus of papers that were using original human annotation or labeling to produce a new training dataset for supervised ML. We restricted our corpus to papers whose classifiers were trained on data from Twitter, for various reasons: First, we did attempt to produce a broader corpus of supervised ML application papers, but found our search queries in academic search engines would either 1) be so broad that most papers were non-applied / theoretical papers or papers re-using public pre-labeled datasets; or 2) that the results were so narrow they excluded many canonical papers in this area, which made us suspect that they were non-representative samples. Sampling to papers using Twitter data has strategic benefits for this kind of initial study. Data from Twitter is of interest to scholars from a variety of disciplines and topical interest areas, in addition to those who have an inherent interest in Twitter as a social media site. As we detail in appendix section \ref{a:keywords}, the papers represented political science, public health, NLP, sentiment analysis, cybersecurity, content moderation, hate speech, information quality, demographic profiling, and more.

We drew the main corpus of ML application papers from ArXiV, the oldest and most established ``preprint'' repositories, originally for researchers to share papers prior to peer review. Today, ArXiV is widely used to share both drafts of papers that have not (yet) passed peer review (``preprints'') and final versions of papers that have passed peer review (often called ``postprints''). Users submit to any number of disciplinary categories and subcategories. Subcategory moderators perform a cursory review to catch spam, blatant hoaxes, and miscategorized papers, but do not review papers for soundness or validity. We sampled all papers published in the Computer Science subcategories of Artificial Intelligence (cs.AI), Machine Learning (cs.LG), Social and Information Networks (cs.SI), Computational Linguistics (cs.CL), Computers and Society (cs.CY), Information Retrieval (cs.IR), and Computer Vision (CS.CV), the Statistics subcategory of Machine Learning (stat.ML), and Social Physics (physics.soc-ph). We filtered for papers in which the title or abstract included at least one of the words ``machine learning'', ``classif*'', or ``supervi*'' (case insensitive). We then filtered to papers in which the title or abstract included at least ``twitter'' or ``tweet'' (case insensitive), which resulted in 494 papers. We used the same query on Elsevier's Scopus database of peer-reviewed articles, selecting 30 randomly sampled articles, which mostly selected from conference proceedings. One paper from the Scopus sample was corrupted, so only 29 papers were examined.

ArXiV is likely not a representative sample of all ML publications. However, we chose it because ArXiV papers are widely accessible to the public, indexed in Google Scholar and other scholarly databases, and are generally considered citeable publications. The fact that many ArXiV papers are not peer-reviewed and that papers posted are not likely representative samples of ML research is worth considering when reflecting on the generalizability of our findings. However, given that such papers are routinely discussed in both academic literature and the popular press means that issues with their reporting of training data is just as crucial. Sampling from ArXiv also lets us examine papers at various stages in the peer-review cycle, breaking out preprints not (yet) published, preprints of later published papers, and postprints of published works. The appendix details both corpora, including an analysis of the topics and fields of papers (in \ref{a:papertypes}), an analysis of the publishers and publication types (e.g. an early preprint of a journal article, a final postprint of a conference proceeding, a preprint never published) and publishers (in \ref{a:pubtypes} and \ref{a:papertypes}). The final dataset can be found on GitHub and Zenodo.\footnote{\url{https://doi.org/10.5281/zenodo.3564844} and \url{https://github.com/staeiou/gigo-fat2020}} 


%

\subsection{Labeling team, training, and workflow}
Our labeling team included one research scientist who led the project (RSG) and undergraduate research assistants, who worked for course credit as part of an university-sponsored research experience program (KY, YY, MD, JQ, RT, and JH). The project began with five students for one semester, four of whom continued on the project for the second semester. A sixth student replaced the student who did not continue. All students had some coursework in computer science and/or data science, with a range of prior experience in machine learning in both a classroom and applied setting. Students' majors and minors included Electrical Engineering \& Computer Science, Data Science, Statistics, and Linguistics.

The labeling workflow was that each week, a set of papers were randomly sampled each week from the unlabled set of 494 ArXiV papers in the corpus. For two weeks, the 30 sampled papers from Scopus were selected. The five students independently reviewed and labeled the same papers each week, using a different web-based spreadsheet to record labels. The team leader synthesized labels and identified disagreement. The team met in person each week to discuss cases of disagreement, working to build a consensus about the proper label (as opposed to purely majority vote). The team leader facilitated these discussions and had the final say when a consensus could not be reached. The papers labeled for the first two weeks were in a training period, in which the team worked on a different set of papers not included in the dataset. In these initial weeks, the team learned the coding schema and the reconciliation process, which were further refined. 

\subsection{Second round verification and reconciliation}

After 164 papers were labeled by five annotators, we conducted a second round of verification. This was necessary both because there were some disagreements in labeling and changes made to the coding schema (discussed in appendix \ref{a:schema}). All labels for all 164 papers were independently re-examined by at least two of the six team members. Annotators were given a summary of the original labels in the first round and were instructed to review all papers, being mindful of how the schema and instructions had changed. We then aggregated, reconciled, and verified labels in the same way as in the first round. For papers where there was no substantive disagreement on any question between those who re-examined it in the second round, the paper's labels were considered to be final. For papers where there was any substantive disagreement on any question, the paper was either discussed to consensus in the same manner as in the first round or decided by the team leader. The final schema and instructions are in the appendix, section \ref{a:instructions}.

Finally, we cleaned up issues with labels around implicit or blank values using rule-based scripts. We learned our process involved some ambiguities around whether a subsequent value needed to be filled in. For example, if a paper was not using crowdworkers, then the instructions for our schema were that the question about crowdworker compensation was to remain blank. However, we found we had cases where ``reported crowdworker compensation'' was ``no'' for papers that did not use crowdworkers. This would be concerning had we had a ``yes'' for such a variable, but found no such cases. We recoded questions about pre-screening for crowdwork platforms (implied by using crowdworkers in original human annotation source) and the number of human annotators.

We measured interrater reliability metrics using mean percent total agreement, or the proportion of cases where all labelers initially gave the same label. This is a more stringent metric than Fleiss's kappa and Krippendorf's alpha, and our data does not fit the assumptions for those widely-used metrics. IRR rates for round one were relatively low: across all questions, the mean percent total agreement was 66.67\%, with the lowest question having a rate of 38.2\%. IRR rates for round two were quite higher: the mean percent total agreement across all questions was 84.80\% and the lowest agreement score was 63.4\% (for ``used external human annotation'', which we discuss later). We are confident about our labeling process, especially because these individual ratings were followed by an expert-adjudicated discussion-based reconciliation process, rather than simply counting majority votes. We detail more information and reflection about interrater reliability in appendix section \ref{a:irr}. 

\subsection{Raw and normalized information scores}
\label{sec:infoscore}
We quantified the information about training data in papers, developing a raw and normalized information score, as different studies demanded different levels of information. For example, our question about whether inter-annotator agreement metrics were reported is only applicable for papers involving multiple annotators. Our questions about whether prescreening was used for crowdwork platforms or whether crowdworker compensation was reported is only relevant for projects using crowdworkers. However, some kinds of information are relevant to all papers that involve original human annotation: who the annotators are (annotation source), annotator training, formal instructions or definitions were given, the number of annotators involved, whether multiple annotators examined the same items, or a link to a publicly-available dataset. 

For raw scores, papers involving original human annotation received one point each for reporting the six items mentioned above. In addition, they received one point per question if they included information for each of the two questions about crowdworkers if the project used crowdworkers, and one point if they reported inter-annotator metrics if the project used multiple annotators per item. For the normalized score, the raw score was divided by the highest possible raw score.\footnote{By 6 if neither crowdworkers nor multiple annotators were used, by 7 if multiple annotators were used, by 8 if crowdworkers were used, and by 9 if both were used.} We only calculated scores for papers involving original human annotation. Finally, we conducted an analysis of information scores by various bibliometric factors, which required determining such factors for all papers. For all ArXiV papers, we determined whether the PDF was a pre-print not (yet) published in another venue, a post-print identical in content to a published version, or a pre-print version of a paper published elsewhere with different content. For all Scopus papers and ArXiV post-prints, we also determined the publisher. We detail these in appendix \ref{a:papertypes}.

\section{Findings}

\subsection{Original classification task}

The first question was whether the paper was conducting an original classification task using supervised machine learning. Our keyword-based process of generating the corpus included many papers not in this scope. However, defining the boundaries of supervised ML and classification tasks is difficult, particularly for papers that are long, complex, and ambiguously worded. We found that some papers claimed to be using ML, but when we examined the details, these did not fall under our definition.  We defined machine learning broadly, using a common working definition in which machine learning includes any automated process that does not exclusively rely on explicit rules, in which the performance of a task increases with additional data. This includes simple linear regressions, for example, and there is much debate about if and when simple linear regressions are a form of ML. However, as we were also looking for classification tasks, linear regressions were only included if it is used to make a prediction in a set of defined classes. We defined an ``original'' classifier to mean a classifier the authors made based on new or old data, which excludes the exclusive use of pre-trained classifiers or models.


\begin{table}[h]
\caption{Original classification task}
\begin{tabular}{lll}
\toprule
{} & Count & Proportion \\
\midrule
Yes    &   142 &     86.59\% \\
No     &    17 &     10.37\% \\
Unsure &     5 &      3.05\% \\
\bottomrule
\end{tabular}
\label{table-original-classification-task}
\end{table}

As table \ref{table-original-classification-task} shows, the overwhelming majority of papers in our dataset were involved in an original classification task. We placed 5 papers in the ``unsure'' category --- meaning they did not give enough detail for us to make this determination, or that they were complex boundary cases. One of the ``unsure'' cases clearly used labels from human annotation, and so we answered the subsequent questions, which is why the counts in Table 2 add up to 143 (as well as some other seeming disparities in later questions).

\subsection{Labels from human annotation}
One of the major issues we had to come to a consensus around was whether a paper used labels from human annotation. We observed a wide range of cases in which human judgment was brought to bear on the curation of training data. Our final definition required that ``the classifier [was] at least in part trained on labeled data that humans made for the purpose of the classification problem.'' We decided on a working definition that excluded many ``clever uses of metadata'' from this category, but did allow some cases of ``self-annotation'' from social media, which were typically the most borderline cases on the other side. For example, one case from our examples we decided was human annotation used specific politically-inflected hashtags to automatically label tweets as for or against a position, for use in stance detection (e.g. \code{\#ProChoice} versus \code{\#ProLife}). However, these cases of self-annotation would all be considered external human annotation rather than original human annotation, and so the subsequent questions about the annotation process would be not applicable. Another set of borderline cases involved papers where no human annotation was involved in the curation of the training dataset that was used to build the classifier, but human annotation was used for validation purposes. We did not consider these to involve human annotation as we originally defined it in our schema, even though the same issues arise with equal significance for the validity of such research. 

\begin{table}[h]
\caption{Labels from human annotation}
\begin{tabular}{lll}
\toprule
{} & Count & Proportion \\
\midrule
Yes    &    93 &     65.04\% \\
No     &    46 &     32.17\% \\
Unsure &     4 &      2.79\% \\
\bottomrule
\end{tabular}
\label{table-labels-from-human-annotation}
\end{table}


\subsection{Used original human annotation and external human annotation}
\label{sec:original_human_annot}

Our next two questions were about whether papers that used human annotation used original human annotation, which we defined as a process in which the paper's authors obtained new labels from humans for items. It is common in ML research to re-use public datasets, and many of papers in our corpus did so. We also found 10 papers in which external and original human annotation was combined to create a new training dataset. For these reasons, we modified our schema to ask separate questions for original and external human annotation data, to capture all three cases (using only original, only external, or both). Tables \ref{table-used-original-human-annotation} and \ref{table-used-external-human-annotation-data} show the breakdown for both questions. We only answered the subsequent questions about the human annotation process for the papers producing an original human annotated dataset.

\begin{table}[h]
\caption{Used original human annotation}
\begin{tabular}{lll}
\toprule
{} & Count & Proportion \\
\midrule
Yes    &    72 &     75.00\% \\
No     &    21 &     21.88\% \\
Unsure &    3  &  3.13\% \\

\bottomrule
\end{tabular}
\label{table-used-original-human-annotation}
\vspace{3px}

\caption{Used external human annotation data}
\begin{tabular}{lll}
\toprule
{} & Count & Proportion \\
\midrule
No     &    61 &     63.54\% \\
Yes    &    32 &     33.33\% \\
Unsure &     3 &      3.13\% \\
\bottomrule
\end{tabular}
\label{table-used-external-human-annotation-data}
\end{table}
\subsection{Original human annotation source}

Our next question asked who the annotators were, for the 74 papers that used original human annotation. The possible options were: the paper's authors, Amazon Mechanical Turk, other crowdworking platforms, experts/professionals, other, and no information. We took phrases like ``we labeled'' (with no other details) to be an implicit declaration that the paper's authors did the labeling. If the paper discussed labelers' qualifications for the task beyond an average person, we labeled it as ``experts / professionals.'' For example, some of our boundary cases involved recruiting students to label sentiment. One study involved labeling tweets with both English and Hindi text and noted that the students were fluent in both languages -- which we considered to be in the ``experts / professionals'' category. Another paper we included in this category recruited students to label tweets with emojis, noting that the recruited students ``are knowledgeable with the context of use of emojis.''

As table \ref{table-original-human-annotation-source} shows, we found a diversity of approaches to the recruitment of human annotators. The plurality of papers involved the paper's authors doing the annotation work themselves. The next highest category was ``no information,'' which was found in almost a quarter of the papers using original human annotation. Experts / professionals was far higher than we expected, although we took any claim of expertise for granted. Crowdworkers constituted a far smaller proportion than we expected, with Amazon Mechanical Turk and other platforms collectively comprising about 15\% of papers. Almost all of the other crowdworking platforms specified were CrowdFlower/FigureEight, with one paper using oDesk. 
\begin{table}[h]
\caption{Original human annotation source}
\begin{tabular}{lll}
\toprule
{} & Count & Proportion \\
\midrule
Paper's authors                  &    22 &     29.73\% \\
No information                   &    18 &     24.32\% \\
Experts / professionals &    16 &     21.62\% \\
Amazon Mechanical Turk                  &     3 &      4.05\% \\
Other crowdwork                  &     8 &     10.81\% \\
Other                            &     7 &      9.46\% \\
\bottomrule
\end{tabular}
\label{table-original-human-annotation-source}
\end{table}

\subsection{Number of human annotators}
\label{sec:num_annot}
Our instructions for the question about the number of human annotators was not precise and had one of the lower levels of inter-rater reliability. If the paper included information about the number of human annotators, the instructions were to put such a number, leaving the field blank for no information. Most of the disagreement was from differences around how papers report the number of annotators used. For example, some papers specified the total number of humans who worked on the project annotating items, while others only specified how many annotators were used per item (particularly for those using crowdworkers), and a few reported both. Some involved a closed set of annotators who all examined the same set of items, similar to how our team operated. Other papers involved an open set of annotators, particularly drawn from crowdworking platforms, but had a consistent number of annotators who reviewed each item. Due to these inconsistencies, we computationally re-coded responses into the presence of information about the number of human annotators. These are both important aspects to discuss, although it is arguably more important to discuss the number of annotators who reviewed each item. In general, having more annotators review each item provides a more robust way of determining the validity of the entire process, although this also requires caluclating inter-annotator agreement metrics.

\begin{table}[h]
\caption{Number of annotators specified}
\begin{tabular}{lll}
\toprule
{} & Count & Proportion \\
\midrule
Yes  &   41 &     55.40\% \\
No &    33 &     44.60\% \\
\bottomrule
\end{tabular}
\label{table-num-of-annotators-specified}
\end{table}

As table \ref{table-num-of-annotators-specified} shows, a slim majority of papers using original human annotation specified the number of annotators involved in some way. Based on our experiences, we typically noticed that papers discussing the number of annotators often fell into two categories: 1) a small closed team (more often 2-3, sometimes 4-6) that were either the papers' authors or recruited directly by the authors, who tended to perform the same amount of work for the duration of the project; or 2) a medium to large (25-500) open set of annotators, typically but not necessarily recruited through a crowdworking platform, who each performed highly variable amounts of work. 
\subsection{Formal definitions and instructions}
Our next question was about whether instructions or guidelines with formal definitions or examples are reportedly given to annotators. Formal definitions and concrete examples are both important, as they help annotators understand how the researchers have operationalized the concept in question and determine edge cases. With no or ambiguous definitions/examples, there could be fundamental misunderstandings that are not captured by inter-annotator agreement metrics, if all annotators make the same misunderstandings. We defined two levels: giving no instructions beyond the text of a question, then giving definitions for each label and/or concrete examples. The paper must describe or refer to instructions given (or include them in supplemental materials), otherwise, we categorized it "No Information". Some borderline cases involved authors labeling the dataset themselves, where the paper presented a formal definition, but only implied that it informed the labeling -- which we took to be a formal definition. As table \ref{table-formal-instructions} shows, the plurality of papers did not provide enough information to make a determination (it is rare for authors to say they did not do something), but 43.2\% provided definitions or examples. 
\begin{table}[h]
\caption{Formal instructions}
\begin{tabular}{lll}
\toprule
{} & Count & Proportion \\
\midrule
No information                                &    35 &     47.30\% \\
Instructions w/ formal definitions/examples &    32 &     43.24\% \\
No instructions beyond question text          &     7 &      9.46\% \\
\bottomrule
\end{tabular}
\label{table-formal-instructions}
\end{table}
\subsection{Training for human annotators}
We defined training for human annotators to involve some kind of interactive process in which the annotators have the opportunity to receive some kind of feedback and/or dialogue about the annotation process. We identified this as a distinct category from both the qualifications of the annotators and the instructions given to annotators, which are examined in other questions. Training typically involved some kind of live session or ongoing meeting in which annotators' progress was evaluated and/or discussed, where annotators had the chance to ask questions or receive feedback on why certain determinations did or did not match definitions or a schema. We used our own team's process as an example of this, and found several papers that used a similar roundtable process, which went into detail about interactions between team members. Cases in which the paper only specified that annotators were given a video or a detailed schema to review were not considered training details, as this was a one-way process and counted as definitions/instructions. 

\begin{table}[h]
\caption{Training for human annotators}
\begin{tabular}{lll}
\toprule
{} & Count & Proportion \\
\midrule
No information        &    63 &     85.14\% \\
Some training details &    11 &     14.86\% \\
\bottomrule
\end{tabular}
\label{table-training-for-human-annotators}
\end{table}

The overwhelming majority of papers did not discuss such issues, as table \ref{table-training-for-human-annotators} shows, with 15\% of papers involving a training session. Because we had a quite strict definition for what constitutes training (versus what many may think of around ``trained annotators''), this is expected. We also are not all that concerned with this low number, as there are many tasks that likely do not require specialized training --- unlike our project, which required both specific expertise in an area and with our complicated schema.
\subsection{Pre-screening for crowdwork platforms}

Crowdwork platforms let employers pre-screen or test for traits, skills, or performance metrics, which significantly narrows the pool of crowdworkers. For example, ``project-specific pre-screening'' involves offering a sample task with known outcomes: if the crowdworker passed, they would be invited to annotate more items. 5 of the 11 papers using crowdworkers reported using this approach. Platforms also often have location-based screening (e.g. US-only), which 2 papers reported using. Some crowdwork platforms have a qualification for workers who have a positive track record based on total employer ratings (e.g. AMT Master). Platforms also offer generic skills-based tests for certain kinds of work (e.g. CrowdFlower's Skill Tests). These last two qualifications were in our coding schema, but no papers reported using them. 

\begin{table}[h]
\caption{Prescreening for crowdwork platforms}
\begin{tabular}{lll}
\toprule
{} & Count & Proportion \\
\midrule
Project-specific prescreening      &     5 &     45.0\% \\
Location qualification             &     2 &     18.0\% \\
No information                     &     4 &     36.0\% \\
\bottomrule
\end{tabular}
\label{table-prescreening-for-crowdwork-platforms}
\end{table}
\subsection{Multiple annotator overlap and reporting inter-annotator agreement}

Our next two questions were about using multiple annotators to review the same items (multiple annotator overlap) and whether inter-annotator agreement metrics were reported. Having multiple independent annotators is typically a foundational best practice in structured content analysis, so that the integrity of the annotations and the schema can be evaluated (although see \cite{McDonald2019}). For multiple annotator overlap, our definitions required papers state whether all or some of the items were labeled by multiple labelers, otherwise ``no information'' was recorded. Then, for papers that did multiple annotator overlap, we examined whether any inter-annotator agreement metric was reported. We did find one paper that did not explicitly state that multiple labelers overlapped, but did report inter-annotator agreement metrics. This implicitly means that at least some of the items were labeled by multiple labelers, but for consistency, we keep the ``no information'' label for this case. We did not record what kind of inter-annotator metric was used, such as Cohen's kappa or Krippendorff's alpha, but many different metrics were used. We also did not record what the exact statistic was, although we did notice a wide variation in what was considered an acceptable or unacceptable score for inter-annotator agreement.

\begin{table}[h]
\caption{Multiple annotator overlap}
\begin{tabular}{lll}
\toprule
{} & Count & Proportion \\
\midrule
No information     &    34 &     45.95\% \\
Yes for all items  &    31 &     41.89\% \\
Yes for some items &     6 &      8.11\% \\
No                 &     3 &      4.05\% \\
\bottomrule
\end{tabular}
\label{table-multiple-annotator-overlap}
\caption{Reported inter-annotator agreement}
\begin{tabular}{lll}
\toprule
{} & Count & Proportion \\
\midrule
Yes &    26 &     70.27\% \\
No  &    11 &     29.73\% \\
\bottomrule
\end{tabular}
\label{table-reported-inter-annotator-agreement}
\end{table}

For multiple annotator overlap, table \ref{table-multiple-annotator-overlap} shows that just under half of all papers that involved an original human annotation task did not provide explicit information one way or the other about whether multiple annotators reviewed each item. This includes the one paper that reported inter-annotator agreement metrics, but did not specify whether overlap was for all items or some items. Only three papers explicitly stated that there was no overlap among annotators, and so it is quite likely that the papers that did not specify such information did not engage in such a practice. For the 37 papers that did involve some kind of multiple annotator overlap, the overwhelming majority of this subsample (84\%) involved multiple annotation of all items, rather than only some items. We also found that for papers that did involve some kind of multiple overlap, the large majority of them (~70\%) did report some metric of inter-annotator agreement, as table \ref{table-reported-inter-annotator-agreement} indicates. 
\subsection{Reported crowdworker compensation}
Crowdworking is often used because of the low cost, which can be far below minimum wage in certain countries. Researchers and crowdworkers have been organizing around  issues related to the exploitation of crowdworkers in research, advocating ethical practices including fair pay \cite{silberman2018responsible}. We examined all papers involving crowdworkers for any indication of compensation, and found zero mentioned compensation. We did find that some papers using other sources of human annotation (e.g. students) discussed compensation for annotators, but this was not in our original schema.

\subsection{Link to dataset available}

Our final question was about whether the paper contained a link to the dataset containing the original human annotated training dataset. Note that this question was only answered for papers involving some kind of original or novel human annotation, and papers that were exclusively re-using an existing open or public dataset were left blank to avoid double-counting.  We did not follow such links or verify that such data was actually available. As table \ref{table-link-to-dataset-available} shows, the overwhelming majority of papers did not include such a link, with 8 papers (10.81\%) using original human-annotated training datasets linking to such data. Given the time, labor, expertise, and funding in creating original human annotated datasets, authors may be hesitant to release such data until they feel they have published as many papers as they can. 

\begin{table}[h]
\caption{Link to dataset available}
\begin{tabular}{lll}
\toprule
{} & Count & Proportion \\
\midrule
No  &    66 &     89.19\% \\
Yes &     8 &     10.81\% \\
\bottomrule
\end{tabular}
\label{table-link-to-dataset-available}
\end{table}
\section{Paper information scores}

The raw and normalized information scores (see section \ref{sec:infoscore} for methodology) were calculated for all papers that involved original human annotation. As previously discussed, our corpora represent a likely non-representative sample of ML research, even if bounded to social computing. Our relatively small sample sizes combined with the number of multiple comparisons would mean that thresholds for statistical significance would need to be quite high. Instead, we present these results to help provide an initial framework and limited results on this issue, intended to help inform a broader and more systematic evaluation the ML literature. We do observe quite varying ranges and distributions of information scores, which does give evidence to the claim that there is substantial and wide variation in the practices around human annotation, training data curation, and research documentation.

\subsection{Overall distributions of information scores}

\begin{figure}[H]
    \centering
    \includegraphics[width=.5\textwidth]{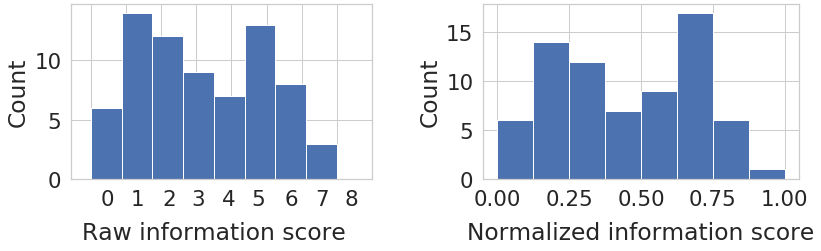}
    \caption{Histograms of raw and normalized information scores for all papers involving original human annotation.}
    \label{fig:info_score_raw_norm_hist}
\end{figure}

Figure \ref{fig:info_score_raw_norm_hist} shows histograms for raw and normalized information scores, which both suggest a bimodal distribution, with fewer papers at the both extremes and the median. This suggests that there are roughly two populations of researchers, with one centered around raw scores of 1-2 and normalized scores of 0.25 and one centered around raw scores of 5 and normalized scores of 0.7. The normalized information score ranged from 0 to 1, with 6 papers having a normalized score of 0 and only 1 paper with a score of 1. The raw information score ranged from 0 to 7, with no paper receiving a full score of 8 or 9, which would have required a study involving crowdworkers, multiple overlap, and open datasets. Overall, the mean normalized information score was 0.441, with a median of 0.429 and a standard deviation of 0.261. The mean raw score was 3.15, with a median of 3.0 and a standard deviation of 2.05.

\subsection{Information scores by corpus and publication type}

\begin{figure}[h]
    \includegraphics[width=.44\textwidth]{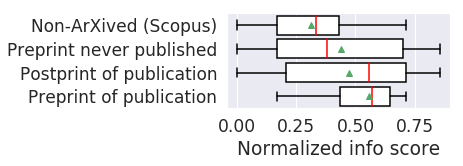}
    \includegraphics[width=.44\textwidth]{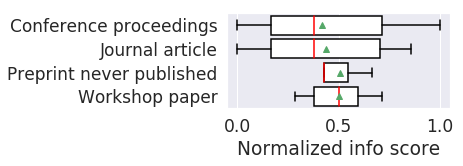}
    \caption{Boxplots of normalized information scores by type of paper. Top: scores by corpus and preprint/postprint status. Bottom: scores from both corpora by publication type.}
    \label{fig:info_score_venue_all}
   \vspace{20px}

\end{figure}

Figure \ref{fig:info_score_venue_all} shows two boxplots\footnote{The main box is the inter-quartile range (IQR), or the 25th \& 75th percentiles. The middle red line is the median, the green triangle is the mean, and the outer whiskers are 5th \& 95th percentiles.} of normalized information scores that are based on different intersecting categories of publication type and status. The left figure compares scores in four categories: all papers in the Scopus sample (non-ArXived), ArXiv preprints that were never (or are not yet) published, and ArXiv preprints that were either postprints or preprints of a traditional publication. The category with the lowest median score are papers from the Scopus sample, which is followed closely by ArXiv preprints never published, although preprints never published had a much larger IQR and standard deviation. Postprints of publications had a similar IQR and standard deviation as preprints never published, but a much higher median score. Preprints of publications had a similar median score as postprints, but with a much smaller IQR and standard deviation. The righthand figure plots publication types for the combined corpora. Conference proceedings and ArXiv preprints never published have somewhat similar medians and IQRs, with journal articles having a higher median of 0.5 and a much narrower IQR. While we hesitate to draw generalizable conclusions, we see these findings indicating a wide range of factors potentially at play.

\subsection{Information scores by publisher}

\begin{figure}[h]
    \centering
    \includegraphics[width=.5\textwidth]{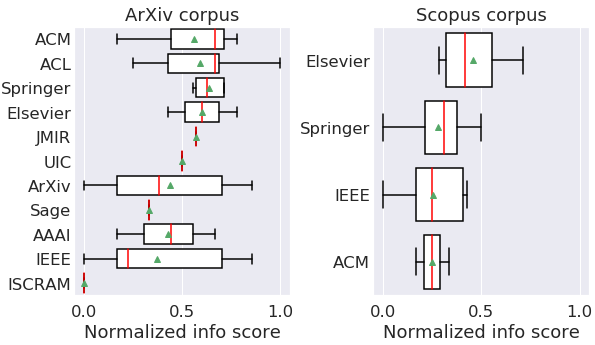}
    \caption{Boxplots of normalized information scores by publisher and corpus, ordered by median score.}
    \label{fig:info_score_pub}
\end{figure}

Figure \ref{fig:info_score_pub} shows boxplots for normalized information scores by publisher, split between papers sampled from  ArXiv and Scopus. The boxplots are ordered by the median score per publisher. In papers in the ArXiv corpus, those that were pre- or post-prints of papers published by the professional societies Association for Computing Machinery (ACM) or Association of Computational Linguistics (ACL) tied for the highest median scores of 0.667, with similar IQRs. These were followed by Springer and Elsevier, with respective medians 0.625 and 0.603 and narrower IQRs. ArXiv preprints not published elsewhere had a median score of 0.381 and the highest IQR and standard deviation (0.289), suggesting that it represents a wide range of papers. The publishers at the lower end of the scale included AAAI, with a median of 0.444 and a narrower IQR, and IEEE, with a median of 0.226 and the second-highest IQR and standard deviation (0.327). Curiously, papers from the Scopus corpus show different results per-publisher, with the median scores of all publishers lower in the Scopus corpus than in the ArXiv corpus. Given the small number of papers in the Scopus sample, we hesitate to draw general conclusions, but suspect it indicates differences between all academic authors and those who post ArXiv postprints.

\section{Concluding discussion}
\subsection{Findings}

In the sample of ML application publications using Twitter data we examined, we found a wide range in levels of documentation about methodological practices in human annotation. While we hesitate to overly generalize our findings to ML at large, these findings do indicate concern, given how crucial the quality of training data is and the difficulty of standardizing human judgment. Yet they also give us hope, as we found a number of papers we considered to be excellent cases of reporting the processes behind their datasets. About half of the papers using original human annotation engaged in some form of multiple overlap, and about 70\% of the papers that did multiple overlap reported metrics of inter-annotator agreement. The distribution of annotation information scores was roughly bimodal, suggesting two distinct populations of those who provide substantially more and less information about training data in their papers. We do see preliminary evidence that papers in our sample published by certain publishers/venues tended to have papers with far more information than others (e.g. ACM and ACL at the top end, followed closely by journal publishers Springer and Elsevier, with IEEE and AAAI proceedings at the lower end). Preprints exclusively published on ArXiv also had the widest range of scores.

\subsection{Implications}

Based on our findings and experiences in this project, we believe human annotation should be considered a core aspect of the research process, with as much attention, care, and concern placed on the annotation process as is currently placed on performance-based metrics like F1 scores. Our findings --- while preliminary, descriptive, and limited in scope --- tell us that there is much room for improvement. This paper also makes steps towards more large-scale and systematic analyses of the research landscape, as well as towards standards and best practices for researchers and reviewers.

Institutions like journals, funders, and disciplinary societies have a major role to play in solutions to these issues. Most publications have strict length maximums, and many papers we scored highly spent a page or more describing their process. Reviewer expectations are crucial in any discussion of the reporting of methodological details in research publications. It could be that some authors did include such details, but were asked to take it out and add other material instead. Authors have incentives to be less open about the messiness inherent in research, as this may open them up to additional criticism. We see many parallels here to issues around reproducibility and open science, which are increasingly being tackled by universal requirements from journals and funders, rather than relying on individuals to change norms. Such research guidelines are common, including the COREQ standard for qualitative data analysis reporting \cite{Tong2007}, a requirement by some journals. A number of proposed standards have been created around datasets for ML \cite{gebru2018datasheets,mitchell2019model,bender2018data,holland2018dataset,barclay2019towards,beretta2018ethical,hind2018increasing}, which are often framed as potential ways to mitigate bias and improve transparency and accountability. Several of these are broader proposals around reporting information about ML classifiers and models, which include various aspects beyond our study. In fact, given the recent explosion of proposals for structured disclosure or transparency documents around ML, the Partnership on AI has recently created the ``ABOUT ML'' working group to arrive at a common format or standard.\footnote{\url{https://www.partnershiponai.org/tag/about-ml/}} \cite{raji_about_2019}

From our perspective, it is important to frame this issue as one of research validity and integrity: what kind of information about training data is needed for researchers, reviewers, and readers to have confidence in the model or classifier? As we observed in our discussions, we became skeptical about papers that did not adequately describe their human annotation processes. However, human annotation is a broad and diverse category of analytical activity, encompassing a wide range of structured human judgment brought to bear on items, some far more straightforward or complex. We saw the wide range papers that were engaged in various forms of annotation or labeling, even though we bounded our study to papers using data from Twitter. One important distinguishing factor is the difficulty of the task and the level of specific knowledge needed to complete it, which can vary significantly. Another key distinction may be between when there is expected to be only one `right' answer and when there might be many valid answers.

Most importantly, we would not want a straightforward checklist to overdetermine issues of model integrity. A number of papers we read were missing details we thought were crucial for understanding that study, but would not make sense for a majority of papers we examined. If a checklist was created, it should not be seen as an end in itself. The classic principle of scientific replicability could be a useful heuristic: does the paper provide enough information about the labeling process such that any reader could (with sufficient resources and access to the same kind of human annotators) conduct a substantively identical human annotation process on their own? We also see a role for technical solutions to help scaffold adherence to these best practices. For example, major qualitative data analysis platforms like MAXQDA or NVivo have built-in support for inter-annotator agreement metrics. Several crowdsourcing and citizen science platforms for data labeling are built to support reconciliation for disagreements. Automated workflow, pipeline, and provenance tracking is an increasing topic in ML, although these can focus more on model building and tuning, taking data as given. We recommend such projects include human annotation as a first-class element, with customization as needed.

Finally, our own experience in this human annotation project studying human annotation projects has shown us the costs and benefits of taking an intensive, detailed, collaborative, and multi-stage approach to human annotation. On one side, we believe that after going through such a long process, we have not only better data, but also a much better contextual understanding of our object of study. Yet on the other hand, even though struggling over the labels and labeling process is an opportunity, our time- and labor-intensive process did have a direct tradeoff with the number of items we were able to annotate. These issues and tradeoffs are important for ML researchers to discuss when designing their own projects and evaluating others. 

\subsection{Limitations and future work}

Our study has limitations, as we only examined a sample of publications in the ML application space. First, we only examined papers that performing a classification task on tweets, which is likely not a representative sample of ML application publications. We would expect to find different results in different domain application areas. Papers in medicine and health may have substantially different practices around reporting training data, due to strict reporting standards in clinical trials and related areas. We also generally examined papers that are posted on ArXiV (in addition to 30 papers sampled from Scopus) and ArXiV is likely to not be a representative sample of academic publications. ArXiV papers are self-submitted and represent a range of publication stages, from drafts not submitted to review, preprints in peer review, and postprints that have passed peer review. Future work should examine different kinds of stratified random samples to examine differences between various publishers, publication types, disciplines, topics, and other factors.

Our study only examined a set of the kinds of issues that scholars and practitioners in ML are examining when they call for greater transparency and accountability through documentation of datasets and models. We have not recorded information about what exactly the rates of inter-annotator agreement are. In particular, we did not record information about the reconciliation or adjudication process for projects which involve multiple overlap (e.g. majority rule, talking to consensus), which we have personally found to be a crucial and difficult process. Other questions we considered but did not include were: the demographics of the labelers, the number of labelers (total and per item), compensation beyond crowdworkers, whether instructions or screenshot of the labeling interface was included, and whether labelers had the option to choose ``unsure'' (vs. being forced to choose a label). We leave this for future work, but also found that each additional question made it more difficult for labelers. We also considered but did not have our team give a holistic score indicating their confidence in the paper (e.g. a 1-5 score, like those used in some peer reviewing processes).

Our study also has limitations that any human annotation project has, and we gained much empathy around the difficulties of human annotation. Our process is not perfect, and as we have analyzed our data, we have identified cases that make us want to change our schema even further or reclassify boundary cases. In future work, we would also recommend using a more structured and constrained system for annotation to capture the text that annotators use to justify their answers to various questions. ML papers are very long and complex, such that our reconciliation and adjudication process was very time-consuming. Finally, we only have access to what the publications say about the work they did, and not the work itself. Future work could improve on this through other methods, such as ethnographic studies of ML practitioners.


\section*{Appendix}
The appendix appears following the references section.
\begin{acks}
This work was funded in part by the Gordon \& Betty Moore Foundation (Grant GBMF3834) and Alfred P. Sloan Foundation (Grant 2013-10-27), as part of the Moore-Sloan Data Science Environments grant to UC-Berkeley. This work was also supported by UC-Berkeley's Undergraduate Research Apprenticeship Program (URAP). We thank many members of UC-Berkeley's Algorithmic Fairness \& Opacity Group (AFOG) for providing invaluable feedback on this project.
 
\end{acks}

\bibliographystyle{ACM-Reference-Format}
\bibliography{refs.bib}


\begin{thebibliography}{68}


\ifx \showCODEN    \undefined \def \showCODEN     #1{\unskip}     \fi
\ifx \showDOI      \undefined \def \showDOI       #1{#1}\fi
\ifx \showISBNx    \undefined \def \showISBNx     #1{\unskip}     \fi
\ifx \showISBNxiii \undefined \def \showISBNxiii  #1{\unskip}     \fi
\ifx \showISSN     \undefined \def \showISSN      #1{\unskip}     \fi
\ifx \showLCCN     \undefined \def \showLCCN      #1{\unskip}     \fi
\ifx \shownote     \undefined \def \shownote      #1{#1}          \fi
\ifx \showarticletitle \undefined \def \showarticletitle #1{#1}   \fi
\ifx \showURL      \undefined \def \showURL       {\relax}        \fi
\providecommand\bibfield[2]{#2}
\providecommand\bibinfo[2]{#2}
\providecommand\natexlab[1]{#1}
\providecommand\showeprint[2][]{arXiv:#2}

\bibitem[\protect\citeauthoryear{Babbage}{Babbage}{1864}]%
        {babbage2011passages}
\bibfield{author}{\bibinfo{person}{Charles Babbage}.}
  \bibinfo{year}{1864}\natexlab{}.
\newblock \bibinfo{booktitle}{\emph{Passages from the Life of a Philosopher}}.
\newblock \bibinfo{publisher}{Longman, Green, Longman, Roberts, and Green},
  \bibinfo{address}{London}.
\newblock


\bibitem[\protect\citeauthoryear{Barclay, Preece, Taylor, and Verma}{Barclay
  et~al\mbox{.}}{2019}]%
        {barclay2019towards}
\bibfield{author}{\bibinfo{person}{Iain Barclay}, \bibinfo{person}{Alun
  Preece}, \bibinfo{person}{Ian Taylor}, {and} \bibinfo{person}{Dinesh Verma}.}
  \bibinfo{year}{2019}\natexlab{}.
\newblock \showarticletitle{Towards Traceability in Data Ecosystems using a
  Bill of Materials Model}.
\newblock \bibinfo{journal}{\emph{arXiv preprint arXiv:1904.04253}}
  (\bibinfo{year}{2019}).
\newblock
\urldef\tempurl%
\url{https://arxiv.org/abs/1904.04253}
\showURL{%
\tempurl}


\bibitem[\protect\citeauthoryear{Bender and Friedman}{Bender and
  Friedman}{2018}]%
        {bender2018data}
\bibfield{author}{\bibinfo{person}{Emily~M Bender} {and} \bibinfo{person}{Batya
  Friedman}.} \bibinfo{year}{2018}\natexlab{}.
\newblock \showarticletitle{Data statements for NLP: Toward mitigating system
  bias and enabling better science}.
\newblock \bibinfo{journal}{\emph{Transactions of the ACL}}
  \bibinfo{volume}{6} (\bibinfo{year}{2018}), \bibinfo{pages}{587--604}.
\newblock
\urldef\tempurl%
\url{https://www.mitpressjournals.org/doi/pdf/10.1162/tacl_a_00041}
\showURL{%
\tempurl}


\bibitem[\protect\citeauthoryear{Beretta, Vetr{\`o}, Lepri, and
  De~Martin}{Beretta et~al\mbox{.}}{2018}]%
        {beretta2018ethical}
\bibfield{author}{\bibinfo{person}{Elena Beretta}, \bibinfo{person}{Antonio
  Vetr{\`o}}, \bibinfo{person}{Bruno Lepri}, {and} \bibinfo{person}{Juan~Carlos
  De~Martin}.} \bibinfo{year}{2018}\natexlab{}.
\newblock \showarticletitle{Ethical and Socially-Aware Data Labels}. In
  \bibinfo{booktitle}{\emph{Annual International Symposium on Information
  Management and Big Data}}. Springer, \bibinfo{pages}{320--327}.
\newblock


\bibitem[\protect\citeauthoryear{Bontcheva, Cunningham, Roberts, Roberts,
  Tablan, Aswani, and Gorrell}{Bontcheva et~al\mbox{.}}{2013}]%
        {bontcheva_gate_2013}
\bibfield{author}{\bibinfo{person}{Kalina Bontcheva}, \bibinfo{person}{Hamish
  Cunningham}, \bibinfo{person}{Ian Roberts}, \bibinfo{person}{Angus Roberts},
  \bibinfo{person}{Valentin Tablan}, \bibinfo{person}{Niraj Aswani}, {and}
  \bibinfo{person}{Genevieve Gorrell}.} \bibinfo{year}{2013}\natexlab{}.
\newblock \showarticletitle{{GATE} {Teamware}: a web-based, collaborative text
  annotation framework}.
\newblock \bibinfo{journal}{\emph{Language Resources and Evaluation}}
  \bibinfo{volume}{47}, \bibinfo{number}{4} (\bibinfo{date}{Dec.}
  \bibinfo{year}{2013}), \bibinfo{pages}{1007--1029}.
\newblock
\showISSN{1574-0218}
\urldef\tempurl%
\url{https://doi.org/10.1007/s10579-013-9215-6}
\showDOI{\tempurl}


\bibitem[\protect\citeauthoryear{Borgman}{Borgman}{2012}]%
        {borgman2012conundrum}
\bibfield{author}{\bibinfo{person}{Christine~L Borgman}.}
  \bibinfo{year}{2012}\natexlab{}.
\newblock \showarticletitle{The conundrum of sharing research data}.
\newblock \bibinfo{journal}{\emph{Journal of the American Society for
  Information Science and Technology}} \bibinfo{volume}{63},
  \bibinfo{number}{6} (\bibinfo{year}{2012}), \bibinfo{pages}{1059--1078}.
\newblock


\bibitem[\protect\citeauthoryear{Bowker and Star}{Bowker and Star}{1999}]%
        {bowker1999sorting}
\bibfield{author}{\bibinfo{person}{Geoffrey~C Bowker} {and}
  \bibinfo{person}{Susan~Leigh Star}.} \bibinfo{year}{1999}\natexlab{}.
\newblock \bibinfo{booktitle}{\emph{Sorting Things Out: Classification and its
  Consequences}}.
\newblock \bibinfo{publisher}{The MIT Press}, \bibinfo{address}{Cambridge, MA}.
\newblock


\bibitem[\protect\citeauthoryear{Burrell}{Burrell}{2016}]%
        {burrell2016machine}
\bibfield{author}{\bibinfo{person}{Jenna Burrell}.}
  \bibinfo{year}{2016}\natexlab{}.
\newblock \showarticletitle{How the machine `thinks': Understanding opacity in
  machine learning algorithms}.
\newblock \bibinfo{journal}{\emph{Big Data \& Society}} \bibinfo{volume}{3},
  \bibinfo{number}{1} (\bibinfo{year}{2016}).
\newblock
\urldef\tempurl%
\url{https://doi.org/10.1177/2053951715622512}
\showURL{%
\tempurl}


\bibitem[\protect\citeauthoryear{Chang, Amershi, and Kamar}{Chang
  et~al\mbox{.}}{2017}]%
        {chang_revolt_2017}
\bibfield{author}{\bibinfo{person}{Joseph~Chee Chang}, \bibinfo{person}{Saleema
  Amershi}, {and} \bibinfo{person}{Ece Kamar}.}
  \bibinfo{year}{2017}\natexlab{}.
\newblock \showarticletitle{Revolt: {Collaborative} {Crowdsourcing} for
  {Labeling} {Machine} {Learning} {Datasets}}. In
  \bibinfo{booktitle}{\emph{Proceedings of the 2017 {CHI} {Conference} on
  {Human} {Factors} in {Computing} {Systems}}} \emph{(\bibinfo{series}{{CHI}
  '17})}. \bibinfo{publisher}{ACM}, \bibinfo{address}{New York, NY, USA},
  \bibinfo{pages}{2334--2346}.
\newblock
\showISBNx{978-1-4503-4655-9}
\urldef\tempurl%
\url{https://doi.org/10.1145/3025453.3026044}
\showDOI{\tempurl}
\newblock
\shownote{event-place: Denver, Colorado, USA.}


\bibitem[\protect\citeauthoryear{Consortium}{Consortium}{2008}]%
        {linguistic2008ace}
\bibfield{author}{\bibinfo{person}{Linguistic~Data Consortium}.}
  \bibinfo{year}{2008}\natexlab{}.
\newblock \bibinfo{title}{ACE (Automatic Content Extraction) English annotation
  guidelines for entities version 6.6}.
\newblock
\newblock
\urldef\tempurl%
\url{https://www.ldc.upenn.edu/sites/www.ldc.upenn.edu/files/english-entities-guidelines-v6.6.pdf}
\showURL{%
\tempurl}


\bibitem[\protect\citeauthoryear{Eubanks}{Eubanks}{2018}]%
        {eubanks2018automating}
\bibfield{author}{\bibinfo{person}{Virginia Eubanks}.}
  \bibinfo{year}{2018}\natexlab{}.
\newblock \bibinfo{booktitle}{\emph{Automating inequality: How high-tech tools
  profile, police, and punish the poor}}.
\newblock \bibinfo{publisher}{St. Martin's Press}.
\newblock


\bibitem[\protect\citeauthoryear{Fecher and Friesike}{Fecher and
  Friesike}{2014}]%
        {fecher_open_2014}
\bibfield{author}{\bibinfo{person}{Benedikt Fecher} {and}
  \bibinfo{person}{Sascha Friesike}.} \bibinfo{year}{2014}\natexlab{}.
\newblock \showarticletitle{Open {Science}: {One} {Term}, {Five} {Schools} of
  {Thought}}.
\newblock In \bibinfo{booktitle}{\emph{Opening {Science}: {The} {Evolving}
  {Guide} on {How} the {Internet} is {Changing} {Research}, {Collaboration} and
  {Scholarly} {Publishing}}}, \bibfield{editor}{\bibinfo{person}{Sönke
  Bartling} {and} \bibinfo{person}{Sascha Friesike}} (Eds.).
  \bibinfo{publisher}{Springer International Publishing},
  \bibinfo{address}{Cham}, \bibinfo{pages}{17--47}.
\newblock
\showISBNx{978-3-319-00026-8}
\urldef\tempurl%
\url{https://doi.org/10.1007/978-3-319-00026-8_2}
\showDOI{\tempurl}


\bibitem[\protect\citeauthoryear{Friedman, Hastie, and Tibshirani}{Friedman
  et~al\mbox{.}}{2009}]%
        {friedman2009elements}
\bibfield{author}{\bibinfo{person}{Jerome Friedman}, \bibinfo{person}{Trevor
  Hastie}, {and} \bibinfo{person}{Robert Tibshirani}.}
  \bibinfo{year}{2009}\natexlab{}.
\newblock \bibinfo{booktitle}{\emph{The Elements of Statistical Learning: Data
  Mining, Inference, and Prediction} (\bibinfo{edition}{2nd} ed.)}.
\newblock \bibinfo{publisher}{Springer}, \bibinfo{address}{New York}.
\newblock


\bibitem[\protect\citeauthoryear{Gebru, Morgenstern, Vecchione, Vaughan,
  Wallach, Daume{\'e}~III, and Crawford}{Gebru et~al\mbox{.}}{2018}]%
        {gebru2018datasheets}
\bibfield{author}{\bibinfo{person}{Timnit Gebru}, \bibinfo{person}{Jamie
  Morgenstern}, \bibinfo{person}{Briana Vecchione},
  \bibinfo{person}{Jennifer~Wortman Vaughan}, \bibinfo{person}{Hanna Wallach},
  \bibinfo{person}{Hal Daume{\'e}~III}, {and} \bibinfo{person}{Kate Crawford}.}
  \bibinfo{year}{2018}\natexlab{}.
\newblock \showarticletitle{Datasheets for Datasets}.
\newblock \bibinfo{journal}{\emph{arXiv preprint arXiv:1803.09010}}
  (\bibinfo{year}{2018}).
\newblock


\bibitem[\protect\citeauthoryear{Gharibi, Walunj, Alanazi, Rella, and
  Lee}{Gharibi et~al\mbox{.}}{2019}]%
        {gharibi_automated_2019}
\bibfield{author}{\bibinfo{person}{Gharib Gharibi}, \bibinfo{person}{Vijay
  Walunj}, \bibinfo{person}{Rakan Alanazi}, \bibinfo{person}{Sirisha Rella},
  {and} \bibinfo{person}{Yugyung Lee}.} \bibinfo{year}{2019}\natexlab{}.
\newblock \showarticletitle{Automated {Management} of {Deep} {Learning}
  {Experiments}}. In \bibinfo{booktitle}{\emph{Proceedings of the 3rd
  {International} {Workshop} on {Data} {Management} for {End}-to-{End}
  {Machine} {Learning}}} \emph{(\bibinfo{series}{{DEEM}'19})}.
  \bibinfo{publisher}{ACM}, \bibinfo{address}{New York, NY, USA},
  \bibinfo{pages}{8:1--8:4}.
\newblock
\showISBNx{978-1-4503-6797-4}
\urldef\tempurl%
\url{https://doi.org/10.1145/3329486.3329495}
\showDOI{\tempurl}
\newblock
\shownote{event-place: Amsterdam, Netherlands.}


\bibitem[\protect\citeauthoryear{Gil, David, Demir, Essawy, Fulweiler, Goodall,
  Karlstrom, Lee, Mills, Oh, Pierce, Pope, Tzeng, Villamizar, and Yu}{Gil
  et~al\mbox{.}}{2016}]%
        {gil_toward_2016}
\bibfield{author}{\bibinfo{person}{Yolanda Gil}, \bibinfo{person}{Cédric~H.
  David}, \bibinfo{person}{Ibrahim Demir}, \bibinfo{person}{Bakinam~T. Essawy},
  \bibinfo{person}{Robinson~W. Fulweiler}, \bibinfo{person}{Jonathan~L.
  Goodall}, \bibinfo{person}{Leif Karlstrom}, \bibinfo{person}{Huikyo Lee},
  \bibinfo{person}{Heath~J. Mills}, \bibinfo{person}{Ji-Hyun Oh},
  \bibinfo{person}{Suzanne~A. Pierce}, \bibinfo{person}{Allen Pope},
  \bibinfo{person}{Mimi~W. Tzeng}, \bibinfo{person}{Sandra~R. Villamizar},
  {and} \bibinfo{person}{Xuan Yu}.} \bibinfo{year}{2016}\natexlab{}.
\newblock \showarticletitle{Toward the {Geoscience} {Paper} of the {Future}:
  {Best} practices for documenting and sharing research from data to software
  to provenance}.
\newblock \bibinfo{journal}{\emph{Earth and Space Science}}
  \bibinfo{volume}{3}, \bibinfo{number}{10} (\bibinfo{year}{2016}),
  \bibinfo{pages}{388--415}.
\newblock
\showISSN{2333-5084}
\urldef\tempurl%
\url{https://doi.org/10.1002/2015EA000136}
\showDOI{\tempurl}


\bibitem[\protect\citeauthoryear{Glaser, Strauss, and Strutzel}{Glaser
  et~al\mbox{.}}{1968}]%
        {glaser}
\bibfield{author}{\bibinfo{person}{Barney~G Glaser}, \bibinfo{person}{Anselm~L
  Strauss}, {and} \bibinfo{person}{Elizabeth Strutzel}.}
  \bibinfo{year}{1968}\natexlab{}.
\newblock \showarticletitle{The discovery of grounded theory; strategies for
  qualitative research}.
\newblock \bibinfo{journal}{\emph{Nursing research}} \bibinfo{volume}{17},
  \bibinfo{number}{4} (\bibinfo{year}{1968}), \bibinfo{pages}{364}.
\newblock


\bibitem[\protect\citeauthoryear{Goodfellow, Bengio, and Courville}{Goodfellow
  et~al\mbox{.}}{2016}]%
        {goodfellow2016deep}
\bibfield{author}{\bibinfo{person}{Ian Goodfellow}, \bibinfo{person}{Yoshua
  Bengio}, {and} \bibinfo{person}{Aaron Courville}.}
  \bibinfo{year}{2016}\natexlab{}.
\newblock \bibinfo{booktitle}{\emph{Deep Learning}}.
\newblock \bibinfo{publisher}{The MIT Press}, \bibinfo{address}{Cambridge, MA}.
\newblock
\newblock
\shownote{\url{http://www.deeplearningbook.org}.}


\bibitem[\protect\citeauthoryear{Goodman, Pepe, Blocker, Borgman, Cranmer,
  Crosas, {Di Stefano}, Gil, Groth, Hedstrom, Hogg, Kashyap, Mahabal,
  Siemiginowska, and Slavkovic}{Goodman et~al\mbox{.}}{2014}]%
        {Goodman2014}
\bibfield{author}{\bibinfo{person}{Alyssa Goodman}, \bibinfo{person}{Alberto
  Pepe}, \bibinfo{person}{Alexander~W. Blocker}, \bibinfo{person}{Christine~L.
  Borgman}, \bibinfo{person}{Kyle Cranmer}, \bibinfo{person}{Merce Crosas},
  \bibinfo{person}{Rosanne {Di Stefano}}, \bibinfo{person}{Yolanda Gil},
  \bibinfo{person}{Paul Groth}, \bibinfo{person}{Margaret Hedstrom},
  \bibinfo{person}{David~W. Hogg}, \bibinfo{person}{Vinay Kashyap},
  \bibinfo{person}{Ashish Mahabal}, \bibinfo{person}{Aneta Siemiginowska},
  {and} \bibinfo{person}{Aleksandra Slavkovic}.}
  \bibinfo{year}{2014}\natexlab{}.
\newblock \showarticletitle{{Ten Simple Rules for the Care and Feeding of
  Scientific Data}}.
\newblock
  \bibinfo{howpublished}{\url{http://dx.plos.org/10.1371/journal.pcbi.1003542}}.
\newblock \bibinfo{journal}{\emph{PLoS Computational Biology}}
  \bibinfo{volume}{10}, \bibinfo{number}{4} (\bibinfo{date}{Apr}
  \bibinfo{year}{2014}), \bibinfo{pages}{e1003542}.
\newblock
\showISSN{1553-7358}
\urldef\tempurl%
\url{https://doi.org/10.1371/journal.pcbi.1003542}
\showDOI{\tempurl}


\bibitem[\protect\citeauthoryear{Goodwin}{Goodwin}{1994}]%
        {Goodwin1994}
\bibfield{author}{\bibinfo{person}{Charles Goodwin}.}
  \bibinfo{year}{1994}\natexlab{}.
\newblock \showarticletitle{{Professional Vision}}.
\newblock \bibinfo{journal}{\emph{American Anthropologist}}
  \bibinfo{volume}{96}, \bibinfo{number}{3} (\bibinfo{date}{sep}
  \bibinfo{year}{1994}), \bibinfo{pages}{606--633}.
\newblock
\showISSN{0002-7294}
\urldef\tempurl%
\url{https://doi.org/10.1525/aa.1994.96.3.02a00100}
\showDOI{\tempurl}


\bibitem[\protect\citeauthoryear{Halfaker and Geiger}{Halfaker and
  Geiger}{2019}]%
        {halfaker2019ores}
\bibfield{author}{\bibinfo{person}{Aaron Halfaker} {and}
  \bibinfo{person}{R~Stuart Geiger}.} \bibinfo{year}{2019}\natexlab{}.
\newblock \showarticletitle{ORES: Lowering Barriers with Participatory Machine
  Learning in Wikipedia}.
\newblock \bibinfo{journal}{\emph{arXiv preprint arXiv:1909.05189}}
  (\bibinfo{year}{2019}).
\newblock
\urldef\tempurl%
\url{https://arxiv.org/pdf/1909.05189.pdf}
\showURL{%
\tempurl}


\bibitem[\protect\citeauthoryear{Hind, Mehta, Mojsilovic, Nair, Ramamurthy,
  Olteanu, and Varshney}{Hind et~al\mbox{.}}{2018}]%
        {hind2018increasing}
\bibfield{author}{\bibinfo{person}{Michael Hind}, \bibinfo{person}{Sameep
  Mehta}, \bibinfo{person}{Aleksandra Mojsilovic}, \bibinfo{person}{Ravi Nair},
  \bibinfo{person}{Karthikeyan~Natesan Ramamurthy}, \bibinfo{person}{Alexandra
  Olteanu}, {and} \bibinfo{person}{Kush~R Varshney}.}
  \bibinfo{year}{2018}\natexlab{}.
\newblock \showarticletitle{Increasing Trust in AI Services through Supplier's
  Declarations of Conformity}.
\newblock \bibinfo{journal}{\emph{arXiv preprint arXiv:1808.07261}}
  (\bibinfo{year}{2018}).
\newblock
\urldef\tempurl%
\url{https://arxiv.org/pdf/1808.07261}
\showURL{%
\tempurl}


\bibitem[\protect\citeauthoryear{Holland, Hosny, Newman, Joseph, and
  Chmielinski}{Holland et~al\mbox{.}}{2018}]%
        {holland2018dataset}
\bibfield{author}{\bibinfo{person}{Sarah Holland}, \bibinfo{person}{Ahmed
  Hosny}, \bibinfo{person}{Sarah Newman}, \bibinfo{person}{Joshua Joseph},
  {and} \bibinfo{person}{Kasia Chmielinski}.} \bibinfo{year}{2018}\natexlab{}.
\newblock \showarticletitle{The dataset nutrition label: A framework to drive
  higher data quality standards}.
\newblock \bibinfo{journal}{\emph{arXiv preprint arXiv:1805.03677}}
  (\bibinfo{year}{2018}).
\newblock
\urldef\tempurl%
\url{https://arxiv.org/abs/1805.03677}
\showURL{%
\tempurl}


\bibitem[\protect\citeauthoryear{Hovy and Lavid}{Hovy and Lavid}{2010}]%
        {hovy2010towards}
\bibfield{author}{\bibinfo{person}{Eduard Hovy} {and} \bibinfo{person}{Julia
  Lavid}.} \bibinfo{year}{2010}\natexlab{}.
\newblock \showarticletitle{Towards a ‘science’ of corpus annotation: a new
  methodological challenge for corpus linguistics}.
\newblock \bibinfo{journal}{\emph{International Journal of Translation}}
  \bibinfo{volume}{22}, \bibinfo{number}{1} (\bibinfo{year}{2010}),
  \bibinfo{pages}{13--36}.
\newblock


\bibitem[\protect\citeauthoryear{Hunter}{Hunter}{2007}]%
        {Matplotlib}
\bibfield{author}{\bibinfo{person}{John~D. Hunter}.}
  \bibinfo{year}{2007}\natexlab{}.
\newblock \showarticletitle{Matplotlib: A 2D Graphics Environment}.
\newblock \bibinfo{journal}{\emph{Computing in Science \& Engineering}}
  \bibinfo{volume}{9}, \bibinfo{number}{3} (\bibinfo{year}{2007}),
  \bibinfo{pages}{90--95}.
\newblock
\urldef\tempurl%
\url{https://doi.org/10.1109/MCSE.2007.55}
\showDOI{\tempurl}
\showeprint{https://aip.scitation.org/doi/pdf/10.1109/MCSE.2007.55}


\bibitem[\protect\citeauthoryear{Jacobs and Wallach}{Jacobs and
  Wallach}{2019}]%
        {jacobs_measurement_2019}
\bibfield{author}{\bibinfo{person}{Abigail~Z. Jacobs} {and}
  \bibinfo{person}{Hanna Wallach}.} \bibinfo{year}{2019}\natexlab{}.
\newblock \showarticletitle{Measurement and {Fairness}}.
\newblock \bibinfo{journal}{\emph{arXiv:1912.05511 [cs]}} (\bibinfo{date}{Dec.}
  \bibinfo{year}{2019}).
\newblock
\urldef\tempurl%
\url{http://arxiv.org/abs/1912.05511}
\showURL{%
\tempurl}
\newblock
\shownote{arXiv: 1912.05511.}


\bibitem[\protect\citeauthoryear{James, Witten, Hastie, and Tibshirani}{James
  et~al\mbox{.}}{2013}]%
        {james2013introduction}
\bibfield{author}{\bibinfo{person}{Gareth James}, \bibinfo{person}{Daniela
  Witten}, \bibinfo{person}{Trevor Hastie}, {and} \bibinfo{person}{Robert
  Tibshirani}.} \bibinfo{year}{2013}\natexlab{}.
\newblock \bibinfo{booktitle}{\emph{An introduction to statistical learning}}.
\newblock \bibinfo{publisher}{Springer}, \bibinfo{address}{New York}.
\newblock


\bibitem[\protect\citeauthoryear{Jones, Oliphant, Peterson,
  et~al\mbox{.}}{Jones et~al\mbox{.}}{2001}]%
        {scipy}
\bibfield{author}{\bibinfo{person}{Eric Jones}, \bibinfo{person}{Travis
  Oliphant}, \bibinfo{person}{Pearu Peterson}, {et~al\mbox{.}}}
  \bibinfo{year}{2001}\natexlab{}.
\newblock \bibinfo{title}{{SciPy}: Open source scientific tools for {Python}}.
\newblock
\newblock
\urldef\tempurl%
\url{http://www.scipy.org/}
\showURL{%
\tempurl}


\bibitem[\protect\citeauthoryear{Kitzes, Turek, and Deniz}{Kitzes
  et~al\mbox{.}}{2018}]%
        {Kitzes2018}
\bibfield{author}{\bibinfo{person}{Justin Kitzes}, \bibinfo{person}{Daniel
  Turek}, {and} \bibinfo{person}{Fatma Deniz}.}
  \bibinfo{year}{2018}\natexlab{}.
\newblock \bibinfo{booktitle}{\emph{{The Practice of Reproducible Research :
  Case Studies and Lessons from the Data-Intensive Sciences}}}.
\newblock \bibinfo{publisher}{University of California Press},
  \bibinfo{address}{Oakland}. 337 pages.
\newblock
\showISBNx{0520294750}
\urldef\tempurl%
\url{http://practicereproducibleresearch.org}
\showURL{%
\tempurl}


\bibitem[\protect\citeauthoryear{Kluyver, Ragan-Kelley, P{\'e}rez, Granger,
  Bussonnier, Frederic, Kelley, Hamrick, Grout, Corlay, Ivanov, Avila, Abdalla,
  and Willing}{Kluyver et~al\mbox{.}}{2016}]%
        {jupyter}
\bibfield{author}{\bibinfo{person}{Thomas Kluyver}, \bibinfo{person}{Benjamin
  Ragan-Kelley}, \bibinfo{person}{Fernando P{\'e}rez}, \bibinfo{person}{Brian
  Granger}, \bibinfo{person}{Matthias Bussonnier}, \bibinfo{person}{Jonathan
  Frederic}, \bibinfo{person}{Kyle Kelley}, \bibinfo{person}{Jessica Hamrick},
  \bibinfo{person}{Jason Grout}, \bibinfo{person}{Sylvain Corlay},
  \bibinfo{person}{Paul Ivanov}, \bibinfo{person}{Dami{\'a}n Avila},
  \bibinfo{person}{Safia Abdalla}, {and} \bibinfo{person}{Carol Willing}.}
  \bibinfo{year}{2016}\natexlab{}.
\newblock \showarticletitle{Jupyter Notebooks: A Publishing format for
  Reproducible Computational Workflows}. In
  \bibinfo{booktitle}{\emph{Positioning and Power in Academic Publishing:
  Players, Agents and Agendas}},
  \bibfield{editor}{\bibinfo{person}{F.~Loizides} {and}
  \bibinfo{person}{B.~Schmidt}} (Eds.). IOS Press,
  \bibinfo{address}{Amsterdam}, \bibinfo{pages}{87 -- 90}.
\newblock
\urldef\tempurl%
\url{https://doi.org/10.3233/978-1-61499-649-1-87}
\showDOI{\tempurl}


\bibitem[\protect\citeauthoryear{Krishnan, Franklin, Goldberg, Wang, and
  Wu}{Krishnan et~al\mbox{.}}{2016}]%
        {krishnan_activeclean_2016}
\bibfield{author}{\bibinfo{person}{Sanjay Krishnan},
  \bibinfo{person}{Michael~J. Franklin}, \bibinfo{person}{Ken Goldberg},
  \bibinfo{person}{Jiannan Wang}, {and} \bibinfo{person}{Eugene Wu}.}
  \bibinfo{year}{2016}\natexlab{}.
\newblock \showarticletitle{{ActiveClean}: {An} {Interactive} {Data} {Cleaning}
  {Framework} {For} {Modern} {Machine} {Learning}}. In
  \bibinfo{booktitle}{\emph{Proceedings of the 2016 {International}
  {Conference} on {Management} of {Data}}} \emph{(\bibinfo{series}{{SIGMOD}
  '16})}. \bibinfo{publisher}{ACM}, \bibinfo{address}{New York, NY, USA},
  \bibinfo{pages}{2117--2120}.
\newblock
\showISBNx{978-1-4503-3531-7}
\urldef\tempurl%
\url{https://doi.org/10.1145/2882903.2899409}
\showDOI{\tempurl}
\newblock
\shownote{event-place: San Francisco, California, USA.}


\bibitem[\protect\citeauthoryear{Latour}{Latour}{1999}]%
        {Latour1999a}
\bibfield{author}{\bibinfo{person}{Bruno Latour}.}
  \bibinfo{year}{1999}\natexlab{}.
\newblock \showarticletitle{{Circulating Reference: Sampling the Soil in the
  Amazon Forest}}.
\newblock In \bibinfo{booktitle}{\emph{Pandora's Hope}}.
  \bibinfo{publisher}{Harvard University Press}, \bibinfo{address}{Cambridge,
  Mass.}
\newblock
\showISBNx{067465336X}


\bibitem[\protect\citeauthoryear{Latour and Woolgar}{Latour and
  Woolgar}{1979}]%
        {latour1979laboratory}
\bibfield{author}{\bibinfo{person}{Bruno Latour} {and} \bibinfo{person}{Steve
  Woolgar}.} \bibinfo{year}{1979}\natexlab{}.
\newblock \bibinfo{booktitle}{\emph{{Laboratory Life: The Social Construction
  of Scientific Facts}}}.
\newblock \bibinfo{publisher}{Sage Publications}, \bibinfo{address}{Beverly
  Hills}.
\newblock


\bibitem[\protect\citeauthoryear{Maeda, Lee, Medero, Medero, Parker, and
  Strassel}{Maeda et~al\mbox{.}}{2008}]%
        {maeda_annotation_2008}
\bibfield{author}{\bibinfo{person}{Kazuaki Maeda}, \bibinfo{person}{Haejoong
  Lee}, \bibinfo{person}{Shawn Medero}, \bibinfo{person}{Julie Medero},
  \bibinfo{person}{Robert Parker}, {and} \bibinfo{person}{Stephanie~M.
  Strassel}.} \bibinfo{year}{2008}\natexlab{}.
\newblock \showarticletitle{Annotation {Tool} {Development} for {Large}-{Scale}
  {Corpus} {Creation} {Projects} at the {Linguistic} {Data} {Consortium}.}. In
  \bibinfo{booktitle}{\emph{Proceedings of the {Sixth} {International}
  {Conference} on {Language} {Resources} and {Evaluation} ({LREC}'08)}},
  Vol.~\bibinfo{volume}{8}.
\newblock
\urldef\tempurl%
\url{http://www.lrec-conf.org/proceedings/lrec2008/pdf/775_paper.pdf}
\showURL{%
\tempurl}


\bibitem[\protect\citeauthoryear{McDonald, Schoenebeck, and Forte}{McDonald
  et~al\mbox{.}}{2019}]%
        {McDonald2019}
\bibfield{author}{\bibinfo{person}{Nora McDonald}, \bibinfo{person}{Sarita
  Schoenebeck}, {and} \bibinfo{person}{Andrea Forte}.}
  \bibinfo{year}{2019}\natexlab{}.
\newblock \showarticletitle{Reliability and Inter-rater Reliability in
  Qualitative Research: Norms and Guidelines for CSCW and HCI Practice}.
\newblock \bibinfo{journal}{\emph{Proc. ACM Hum.-Comput. Interact.}}
  \bibinfo{volume}{3}, \bibinfo{number}{CSCW}, Article \bibinfo{articleno}{72}
  (\bibinfo{date}{Nov.} \bibinfo{year}{2019}), \bibinfo{numpages}{23}~pages.
\newblock
\showISSN{2573-0142}
\urldef\tempurl%
\url{https://doi.org/10.1145/3359174}
\showDOI{\tempurl}


\bibitem[\protect\citeauthoryear{McKinney}{McKinney}{2010}]%
        {pandas}
\bibfield{author}{\bibinfo{person}{Wes McKinney}.}
  \bibinfo{year}{2010}\natexlab{}.
\newblock \showarticletitle{{Data Structures for Statistical Computing in
  Python}}. In \bibinfo{booktitle}{\emph{Proceedings of the 9th Python in
  Science Conference}}, \bibfield{editor}{\bibinfo{person}{St{\'{e}}fan van~der
  Walt} {and} \bibinfo{person}{Jarrod Millman}} (Eds.).
  \bibinfo{pages}{51--56}.
\newblock
\urldef\tempurl%
\url{http://conference.scipy.org/proceedings/scipy2010/mckinney.html}
\showURL{%
\tempurl}


\bibitem[\protect\citeauthoryear{Medeiros and Ball}{Medeiros and Ball}{2017}]%
        {Medeiros2017}
\bibfield{author}{\bibinfo{person}{N. Medeiros} {and} \bibinfo{person}{R.J.
  Ball}.} \bibinfo{year}{2017}\natexlab{}.
\newblock \showarticletitle{Teaching Integrity in Empirical Economics: The
  Pedagogy of Reproducible Science in Undergraduate Education}.
\newblock In \bibinfo{booktitle}{\emph{Undergraduate Research and the Academic
  Librarian: Case Studies and Best Practices}},
  \bibfield{editor}{\bibinfo{person}{M.K. Hensley} {and}
  \bibinfo{person}{S.~Davis-Kahl}} (Eds.). \bibinfo{publisher}{Association of
  College \& Research Libraries}, \bibinfo{address}{Chicago}.
\newblock
\urldef\tempurl%
\url{https://scholarship.haverford.edu/cgi/viewcontent.cgi?article=1189}
\showURL{%
\tempurl}


\bibitem[\protect\citeauthoryear{Mellin}{Mellin}{1957}]%
        {mellin1957work}
\bibfield{author}{\bibinfo{person}{WD Mellin}.}
  \bibinfo{year}{1957}\natexlab{}.
\newblock \showarticletitle{Work with new electronic `brains' opens field for
  army math experts}.
\newblock \bibinfo{journal}{\emph{The Hammond Times}}  \bibinfo{volume}{10}
  (\bibinfo{year}{1957}), \bibinfo{pages}{66}.
\newblock


\bibitem[\protect\citeauthoryear{Mitchell, Wu, Zaldivar, Barnes, Vasserman,
  Hutchinson, Spitzer, Raji, and Gebru}{Mitchell et~al\mbox{.}}{2019}]%
        {mitchell2019model}
\bibfield{author}{\bibinfo{person}{Margaret Mitchell}, \bibinfo{person}{Simone
  Wu}, \bibinfo{person}{Andrew Zaldivar}, \bibinfo{person}{Parker Barnes},
  \bibinfo{person}{Lucy Vasserman}, \bibinfo{person}{Ben Hutchinson},
  \bibinfo{person}{Elena Spitzer}, \bibinfo{person}{Inioluwa~Deborah Raji},
  {and} \bibinfo{person}{Timnit Gebru}.} \bibinfo{year}{2019}\natexlab{}.
\newblock \showarticletitle{Model cards for model reporting}. In
  \bibinfo{booktitle}{\emph{Proceedings of the Conference on Fairness,
  Accountability, and Transparency}}. ACM, \bibinfo{pages}{220--229}.
\newblock


\bibitem[\protect\citeauthoryear{Mozeti{\v{c}}, Gr{\v{c}}ar, and
  Smailovi{\'{c}}}{Mozeti{\v{c}} et~al\mbox{.}}{2016}]%
        {Mozetic2016}
\bibfield{author}{\bibinfo{person}{Igor Mozeti{\v{c}}}, \bibinfo{person}{Miha
  Gr{\v{c}}ar}, {and} \bibinfo{person}{Jasmina Smailovi{\'{c}}}.}
  \bibinfo{year}{2016}\natexlab{}.
\newblock \showarticletitle{{Multilingual Twitter Sentiment Classification: The
  Role of Human Annotators}}.
\newblock \bibinfo{journal}{\emph{PLOS ONE}} \bibinfo{volume}{11},
  \bibinfo{number}{5} (\bibinfo{date}{may} \bibinfo{year}{2016}),
  \bibinfo{pages}{e0155036}.
\newblock
\showISSN{1932-6203}
\urldef\tempurl%
\url{https://doi.org/10.1371/journal.pone.0155036}
\showDOI{\tempurl}


\bibitem[\protect\citeauthoryear{Nakayama, Kubo, Kamura, Taniguchi, and
  Liang}{Nakayama et~al\mbox{.}}{2018}]%
        {doccano}
\bibfield{author}{\bibinfo{person}{Hiroki Nakayama}, \bibinfo{person}{Takahiro
  Kubo}, \bibinfo{person}{Junya Kamura}, \bibinfo{person}{Yasufumi Taniguchi},
  {and} \bibinfo{person}{Xu Liang}.} \bibinfo{year}{2018}\natexlab{}.
\newblock \bibinfo{title}{{doccano}: Text Annotation Tool for Human}.
\newblock
\newblock
\urldef\tempurl%
\url{https://github.com/doccano/doccano}
\showURL{%
\tempurl}
\newblock
\shownote{Software available from https://github.com/doccano/doccano.}


\bibitem[\protect\citeauthoryear{Nelson}{Nelson}{2017}]%
        {nelson}
\bibfield{author}{\bibinfo{person}{Laura~K Nelson}.}
  \bibinfo{year}{2017}\natexlab{}.
\newblock \showarticletitle{Computational grounded theory: A methodological
  framework}.
\newblock \bibinfo{journal}{\emph{Sociological Methods \& Research}}
  (\bibinfo{year}{2017}).
\newblock


\bibitem[\protect\citeauthoryear{Oleinik, Popova, Kirdina, and
  Shatalova}{Oleinik et~al\mbox{.}}{2014}]%
        {oleinik_choice_2014}
\bibfield{author}{\bibinfo{person}{Anton Oleinik}, \bibinfo{person}{Irina
  Popova}, \bibinfo{person}{Svetlana Kirdina}, {and} \bibinfo{person}{Tatyana
  Shatalova}.} \bibinfo{year}{2014}\natexlab{}.
\newblock \showarticletitle{On the choice of measures of reliability and
  validity in the content-analysis of texts}.
\newblock \bibinfo{journal}{\emph{Quality \& Quantity}} \bibinfo{volume}{48},
  \bibinfo{number}{5} (\bibinfo{date}{Sept.} \bibinfo{year}{2014}),
  \bibinfo{pages}{2703--2718}.
\newblock
\showISSN{1573-7845}
\urldef\tempurl%
\url{https://doi.org/10.1007/s11135-013-9919-0}
\showDOI{\tempurl}


\bibitem[\protect\citeauthoryear{Pasquale}{Pasquale}{2015}]%
        {pasquale2015black}
\bibfield{author}{\bibinfo{person}{Frank Pasquale}.}
  \bibinfo{year}{2015}\natexlab{}.
\newblock \bibinfo{booktitle}{\emph{{The Black Box Society: The Secret
  Algorithms That Control Money and Information}}}.
\newblock \bibinfo{publisher}{Harvard University Press},
  \bibinfo{address}{Cambridge}.
\newblock


\bibitem[\protect\citeauthoryear{P\'erez and Granger}{P\'erez and
  Granger}{2007}]%
        {ipython}
\bibfield{author}{\bibinfo{person}{Fernando P\'erez} {and}
  \bibinfo{person}{Brian~E. Granger}.} \bibinfo{year}{2007}\natexlab{}.
\newblock \showarticletitle{{IP}ython: a System for Interactive Scientific
  Computing}.
\newblock \bibinfo{journal}{\emph{Computing in Science and Engineering}}
  \bibinfo{volume}{9}, \bibinfo{number}{3} (\bibinfo{date}{May}
  \bibinfo{year}{2007}), \bibinfo{pages}{21--29}.
\newblock
\showISSN{1521-9615}
\urldef\tempurl%
\url{https://doi.org/10.1109/MCSE.2007.53}
\showDOI{\tempurl}


\bibitem[\protect\citeauthoryear{{P}roject {J}upyter, {M}atthias {B}ussonnier,
  {J}essica {F}orde, {J}eremy {F}reeman, {B}rian {G}ranger, {T}im {H}ead,
  {C}hris {H}oldgraf, {K}yle {K}elley, {G}ladys {N}alvarte, {A}ndrew
  {O}sheroff, {P}acer, {Y}uvi {P}anda, {F}ernando {P}erez, {B}enjamin~{R}agan
  {K}elley, and {C}arol {W}illing}{{P}roject {J}upyter et~al\mbox{.}}{2018}]%
        {binder}
\bibfield{author}{\bibinfo{person}{{P}roject {J}upyter},
  \bibinfo{person}{{M}atthias {B}ussonnier}, \bibinfo{person}{{J}essica
  {F}orde}, \bibinfo{person}{{J}eremy {F}reeman}, \bibinfo{person}{{B}rian
  {G}ranger}, \bibinfo{person}{{T}im {H}ead}, \bibinfo{person}{{C}hris
  {H}oldgraf}, \bibinfo{person}{{K}yle {K}elley}, \bibinfo{person}{{G}ladys
  {N}alvarte}, \bibinfo{person}{{A}ndrew {O}sheroff}, \bibinfo{person}{{M}
  {P}acer}, \bibinfo{person}{{Y}uvi {P}anda}, \bibinfo{person}{{F}ernando
  {P}erez}, \bibinfo{person}{{B}enjamin~{R}agan {K}elley}, {and}
  \bibinfo{person}{{C}arol {W}illing}.} \bibinfo{year}{2018}\natexlab{}.
\newblock \showarticletitle{{B}inder 2.0 - {R}eproducible, Interactive,
  Sharable Environments for Science at Scale}. In
  \bibinfo{booktitle}{\emph{{P}roceedings of the 17th {P}ython in {S}cience
  {C}onference}}, \bibfield{editor}{\bibinfo{person}{{F}atih {A}kici},
  \bibinfo{person}{{D}avid {L}ippa}, \bibinfo{person}{{D}illon {N}iederhut},
  {and} \bibinfo{person}{{M} {P}acer}} (Eds.). \bibinfo{pages}{113 -- 120}.
\newblock
\urldef\tempurl%
\url{https://doi.org/10.25080/Majora-4af1f417-011}
\showDOI{\tempurl}


\bibitem[\protect\citeauthoryear{Pérez-Pérez, Glez-Peña, Fdez-Riverola, and
  Lourenço}{Pérez-Pérez et~al\mbox{.}}{2015}]%
        {perez_marky_2015}
\bibfield{author}{\bibinfo{person}{Martín Pérez-Pérez},
  \bibinfo{person}{Daniel Glez-Peña}, \bibinfo{person}{Florentino
  Fdez-Riverola}, {and} \bibinfo{person}{Anália Lourenço}.}
  \bibinfo{year}{2015}\natexlab{}.
\newblock \showarticletitle{Marky: {A} tool supporting annotation consistency
  in multi-user and iterative document annotation projects}.
\newblock \bibinfo{journal}{\emph{Computer Methods and Programs in
  Biomedicine}} \bibinfo{volume}{118}, \bibinfo{number}{2}
  (\bibinfo{date}{Feb.} \bibinfo{year}{2015}), \bibinfo{pages}{242--251}.
\newblock
\showISSN{0169-2607}
\urldef\tempurl%
\url{https://doi.org/10.1016/j.cmpb.2014.11.005}
\showDOI{\tempurl}


\bibitem[\protect\citeauthoryear{Quarfoot and Levine}{Quarfoot and
  Levine}{2016}]%
        {quarfoot_how_2016}
\bibfield{author}{\bibinfo{person}{David Quarfoot} {and}
  \bibinfo{person}{Richard~A. Levine}.} \bibinfo{year}{2016}\natexlab{}.
\newblock \showarticletitle{How {Robust} {Are} {Multirater} {Interrater}
  {Reliability} {Indices} to {Changes} in {Frequency} {Distribution}?}
\newblock \bibinfo{journal}{\emph{The American Statistician}}
  \bibinfo{volume}{70}, \bibinfo{number}{4} (\bibinfo{date}{Oct.}
  \bibinfo{year}{2016}), \bibinfo{pages}{373--384}.
\newblock
\showISSN{0003-1305}
\urldef\tempurl%
\url{https://doi.org/10.1080/00031305.2016.1141708}
\showDOI{\tempurl}


\bibitem[\protect\citeauthoryear{Raji and Yang}{Raji and Yang}{2019}]%
        {raji_about_2019}
\bibfield{author}{\bibinfo{person}{Inioluwa~Deborah Raji} {and}
  \bibinfo{person}{Jingying Yang}.} \bibinfo{year}{2019}\natexlab{}.
\newblock \showarticletitle{{ABOUT} {ML}: {Annotation} and {Benchmarking} on
  {Understanding} and {Transparency} of {Machine} {Learning} {Lifecycles}}.
\newblock \bibinfo{journal}{\emph{arXiv:1912.06166 [cs, stat]}}
  (\bibinfo{date}{Dec.} \bibinfo{year}{2019}).
\newblock
\urldef\tempurl%
\url{http://arxiv.org/abs/1912.06166}
\showURL{%
\tempurl}
\newblock
\shownote{arXiv: 1912.06166.}


\bibitem[\protect\citeauthoryear{Raykar and Yu}{Raykar and Yu}{2012}]%
        {raykar2012eliminating}
\bibfield{author}{\bibinfo{person}{Vikas~C Raykar} {and}
  \bibinfo{person}{Shipeng Yu}.} \bibinfo{year}{2012}\natexlab{}.
\newblock \showarticletitle{Eliminating spammers and ranking annotators for
  crowdsourced labeling tasks}.
\newblock \bibinfo{journal}{\emph{Journal of Machine Learning Research}}
  \bibinfo{volume}{13}, \bibinfo{number}{Feb} (\bibinfo{year}{2012}),
  \bibinfo{pages}{491--518}.
\newblock


\bibitem[\protect\citeauthoryear{Riff, Lacy, and Fico}{Riff
  et~al\mbox{.}}{2013}]%
        {riff2013analyzing}
\bibfield{author}{\bibinfo{person}{Daniel Riff}, \bibinfo{person}{Stephen
  Lacy}, {and} \bibinfo{person}{Frederick Fico}.}
  \bibinfo{year}{2013}\natexlab{}.
\newblock \bibinfo{booktitle}{\emph{Analyzing media messages: Using
  quantitative content analysis in research}}.
\newblock \bibinfo{publisher}{Routledge}, \bibinfo{address}{New York}.
\newblock


\bibitem[\protect\citeauthoryear{Sabou, Bontcheva, Derczynski, and
  Scharl}{Sabou et~al\mbox{.}}{2014}]%
        {sabouetal2014}
\bibfield{author}{\bibinfo{person}{Marta Sabou}, \bibinfo{person}{Kalina
  Bontcheva}, \bibinfo{person}{Leon Derczynski}, {and} \bibinfo{person}{Arno
  Scharl}.} \bibinfo{year}{2014}\natexlab{}.
\newblock \showarticletitle{Corpus Annotation through Crowdsourcing: Towards
  Best Practice Guidelines}. In \bibinfo{booktitle}{\emph{Proceedings of the
  Ninth International Conference on Language Resources and Evaluation
  ({LREC}'14)}}. \bibinfo{publisher}{European Language Resources Association
  (ELRA)}, \bibinfo{address}{Reykjavik, Iceland}, \bibinfo{pages}{859--866}.
\newblock
\urldef\tempurl%
\url{http://www.lrec-conf.org/proceedings/lrec2014/pdf/497_Paper.pdf}
\showURL{%
\tempurl}


\bibitem[\protect\citeauthoryear{Sallans and Donnelly}{Sallans and
  Donnelly}{2012}]%
        {sallans2012dmp}
\bibfield{author}{\bibinfo{person}{Andrew Sallans} {and}
  \bibinfo{person}{Martin Donnelly}.} \bibinfo{year}{2012}\natexlab{}.
\newblock \showarticletitle{DMP Online and DMPTool: Different Strategies
  Towards a Shared Goal}.
\newblock \bibinfo{journal}{\emph{International Journal of Digital Curation}}
  \bibinfo{volume}{7}, \bibinfo{number}{2} (\bibinfo{year}{2012}),
  \bibinfo{pages}{123--129}.
\newblock
\urldef\tempurl%
\url{https://doi.org/10.2218/ijdc.v7i2.235}
\showURL{%
\tempurl}


\bibitem[\protect\citeauthoryear{Schelter, Böse, Kirschnick, Klein, and
  Seufert}{Schelter et~al\mbox{.}}{2017}]%
        {schelter_automatically_2017}
\bibfield{author}{\bibinfo{person}{Sebastian Schelter},
  \bibinfo{person}{Joos-Hendrik Böse}, \bibinfo{person}{Johannes Kirschnick},
  \bibinfo{person}{Thoralf Klein}, {and} \bibinfo{person}{Stephan Seufert}.}
  \bibinfo{year}{2017}\natexlab{}.
\newblock \showarticletitle{Automatically tracking metadata and provenance of
  machine learning experiments}. In \bibinfo{booktitle}{\emph{Machine
  {Learning} {Systems} workshop at {NIPS}}}.
\newblock


\bibitem[\protect\citeauthoryear{Schelter, Lange, Schmidt, Celikel, Biessmann,
  and Grafberger}{Schelter et~al\mbox{.}}{2018}]%
        {schelter_automating_2018}
\bibfield{author}{\bibinfo{person}{Sebastian Schelter}, \bibinfo{person}{Dustin
  Lange}, \bibinfo{person}{Philipp Schmidt}, \bibinfo{person}{Meltem Celikel},
  \bibinfo{person}{Felix Biessmann}, {and} \bibinfo{person}{Andreas
  Grafberger}.} \bibinfo{year}{2018}\natexlab{}.
\newblock \showarticletitle{Automating {Large}-scale {Data} {Quality}
  {Verification}}.
\newblock \bibinfo{journal}{\emph{Proc. VLDB Endow.}} \bibinfo{volume}{11},
  \bibinfo{number}{12} (\bibinfo{date}{Aug.} \bibinfo{year}{2018}),
  \bibinfo{pages}{1781--1794}.
\newblock
\showISSN{2150-8097}
\urldef\tempurl%
\url{https://doi.org/10.14778/3229863.3229867}
\showDOI{\tempurl}


\bibitem[\protect\citeauthoryear{Schreier, Wilson, and Resnik}{Schreier
  et~al\mbox{.}}{2006}]%
        {schreier2006academic}
\bibfield{author}{\bibinfo{person}{Alan~A Schreier}, \bibinfo{person}{Kenneth
  Wilson}, {and} \bibinfo{person}{David Resnik}.}
  \bibinfo{year}{2006}\natexlab{}.
\newblock \showarticletitle{Academic research record-keeping: Best practices
  for individuals, group leaders, and institutions}.
\newblock \bibinfo{journal}{\emph{Academic medicine: journal of the Association
  of American Medical Colleges}} \bibinfo{volume}{81}, \bibinfo{number}{1}
  (\bibinfo{year}{2006}), \bibinfo{pages}{42}.
\newblock
\urldef\tempurl%
\url{https://www.ncbi.nlm.nih.gov/pmc/articles/PMC3943904/}
\showURL{%
\tempurl}


\bibitem[\protect\citeauthoryear{Scott}{Scott}{1998}]%
        {scott_seeing_1998}
\bibfield{author}{\bibinfo{person}{James~C. Scott}.}
  \bibinfo{year}{1998}\natexlab{}.
\newblock \bibinfo{booktitle}{\emph{Seeing like a state: {How} certain schemes
  to improve the human condition have failed}}.
\newblock \bibinfo{publisher}{Yale University Press}.
\newblock


\bibitem[\protect\citeauthoryear{Silberman, Tomlinson, LaPlante, Ross, Irani,
  and Zaldivar}{Silberman et~al\mbox{.}}{2018}]%
        {silberman2018responsible}
\bibfield{author}{\bibinfo{person}{M~Six Silberman}, \bibinfo{person}{Bill
  Tomlinson}, \bibinfo{person}{Rochelle LaPlante}, \bibinfo{person}{Joel Ross},
  \bibinfo{person}{Lilly Irani}, {and} \bibinfo{person}{Andrew Zaldivar}.}
  \bibinfo{year}{2018}\natexlab{}.
\newblock \showarticletitle{Responsible research with crowds: pay crowdworkers
  at least minimum wage.}
\newblock \bibinfo{journal}{\emph{Commun. ACM}} \bibinfo{volume}{61},
  \bibinfo{number}{3} (\bibinfo{year}{2018}), \bibinfo{pages}{39--41}.
\newblock


\bibitem[\protect\citeauthoryear{Simpson, Page, and De~Roure}{Simpson
  et~al\mbox{.}}{2014}]%
        {Simpson2014}
\bibfield{author}{\bibinfo{person}{Robert Simpson}, \bibinfo{person}{Kevin~R.
  Page}, {and} \bibinfo{person}{David De~Roure}.}
  \bibinfo{year}{2014}\natexlab{}.
\newblock \showarticletitle{Zooniverse: Observing the World's Largest Citizen
  Science Platform}. In \bibinfo{booktitle}{\emph{Proceedings of the 23rd
  International Conference on World Wide Web}} \emph{(\bibinfo{series}{WWW '14
  Companion})}. \bibinfo{publisher}{ACM}, \bibinfo{address}{New York, NY, USA},
  \bibinfo{pages}{1049--1054}.
\newblock
\showISBNx{978-1-4503-2745-9}
\urldef\tempurl%
\url{https://doi.org/10.1145/2567948.2579215}
\showDOI{\tempurl}


\bibitem[\protect\citeauthoryear{Singh, Cobbe, and Norval}{Singh
  et~al\mbox{.}}{2019}]%
        {singh_decision_2019}
\bibfield{author}{\bibinfo{person}{Jatinder Singh}, \bibinfo{person}{Jennifer
  Cobbe}, {and} \bibinfo{person}{Chris Norval}.}
  \bibinfo{year}{2019}\natexlab{}.
\newblock \showarticletitle{Decision {Provenance}: {Harnessing} {Data} {Flow}
  for {Accountable} {Systems}}.
\newblock \bibinfo{journal}{\emph{IEEE Access}}  \bibinfo{volume}{7}
  (\bibinfo{year}{2019}), \bibinfo{pages}{6562--6574}.
\newblock
\showISSN{2169-3536}
\urldef\tempurl%
\url{https://doi.org/10.1109/ACCESS.2018.2887201}
\showDOI{\tempurl}


\bibitem[\protect\citeauthoryear{Sober{\'o}n, Aroyo, Welty, Inel, Lin, and
  Overmeen}{Sober{\'o}n et~al\mbox{.}}{2013}]%
        {soberon2013measuring}
\bibfield{author}{\bibinfo{person}{Guillermo Sober{\'o}n},
  \bibinfo{person}{Lora Aroyo}, \bibinfo{person}{Chris Welty},
  \bibinfo{person}{Oana Inel}, \bibinfo{person}{Hui Lin}, {and}
  \bibinfo{person}{Manfred Overmeen}.} \bibinfo{year}{2013}\natexlab{}.
\newblock \showarticletitle{Measuring crowd truth: Disagreement metrics
  combined with worker behavior filters}. In \bibinfo{booktitle}{\emph{CrowdSem
  2013 Workshop}}.
\newblock


\bibitem[\protect\citeauthoryear{Stuart}{Stuart}{2004}]%
        {stuart2004databases}
\bibfield{author}{\bibinfo{person}{Guy Stuart}.}
  \bibinfo{year}{2004}\natexlab{}.
\newblock \showarticletitle{{Databases, Felons, and Voting: Bias and
  Partisanship of the Florida Felons List in the 2000 Elections}}.
\newblock \bibinfo{journal}{\emph{Political Science Quarterly}}
  \bibinfo{volume}{119}, \bibinfo{number}{3} (\bibinfo{date}{sep}
  \bibinfo{year}{2004}), \bibinfo{pages}{453--475}.
\newblock
\showISSN{00323195}
\urldef\tempurl%
\url{https://doi.org/10.2307/20202391}
\showDOI{\tempurl}


\bibitem[\protect\citeauthoryear{Tong, Sainsbury, and Craig}{Tong
  et~al\mbox{.}}{2007}]%
        {Tong2007}
\bibfield{author}{\bibinfo{person}{A. Tong}, \bibinfo{person}{P. Sainsbury},
  {and} \bibinfo{person}{J. Craig}.} \bibinfo{year}{2007}\natexlab{}.
\newblock \showarticletitle{{Consolidated criteria for reporting qualitative
  research (COREQ): a 32-item checklist for interviews and focus groups}}.
\newblock \bibinfo{journal}{\emph{International Journal for Quality in Health
  Care}} \bibinfo{volume}{19}, \bibinfo{number}{6} (\bibinfo{date}{sep}
  \bibinfo{year}{2007}), \bibinfo{pages}{349--357}.
\newblock
\showISSN{1353-4505}
\urldef\tempurl%
\url{https://doi.org/10.1093/intqhc/mzm042}
\showDOI{\tempurl}


\bibitem[\protect\citeauthoryear{van~der Walt, Colbert, and Varoquaux}{van~der
  Walt et~al\mbox{.}}{2011}]%
        {numpy}
\bibfield{author}{\bibinfo{person}{S. van~der Walt}, \bibinfo{person}{S.~C.
  Colbert}, {and} \bibinfo{person}{G. Varoquaux}.}
  \bibinfo{year}{2011}\natexlab{}.
\newblock \showarticletitle{The NumPy Array: A Structure for Efficient
  Numerical Computation}.
\newblock \bibinfo{journal}{\emph{Computing in Science Engineering}}
  \bibinfo{volume}{13}, \bibinfo{number}{2} (\bibinfo{date}{March}
  \bibinfo{year}{2011}), \bibinfo{pages}{22--30}.
\newblock
\showISSN{1521-9615}
\urldef\tempurl%
\url{https://doi.org/10.1109/MCSE.2011.37}
\showDOI{\tempurl}


\bibitem[\protect\citeauthoryear{van Rossum}{van Rossum}{1995}]%
        {python}
\bibfield{author}{\bibinfo{person}{Guido van Rossum}.}
  \bibinfo{year}{1995}\natexlab{}.
\newblock \bibinfo{title}{Python Library Reference}.
\newblock
\newblock
\urldef\tempurl%
\url{https://ir.cwi.nl/pub/5009/05009D.pdf}
\showURL{%
\tempurl}


\bibitem[\protect\citeauthoryear{Von~Ahn, Maurer, McMillen, Abraham, and
  Blum}{Von~Ahn et~al\mbox{.}}{2008}]%
        {von2008recaptcha}
\bibfield{author}{\bibinfo{person}{Luis Von~Ahn}, \bibinfo{person}{Benjamin
  Maurer}, \bibinfo{person}{Colin McMillen}, \bibinfo{person}{David Abraham},
  {and} \bibinfo{person}{Manuel Blum}.} \bibinfo{year}{2008}\natexlab{}.
\newblock \showarticletitle{recaptcha: Human-based character recognition via
  web security measures}.
\newblock \bibinfo{journal}{\emph{Science}} \bibinfo{volume}{321},
  \bibinfo{number}{5895} (\bibinfo{year}{2008}), \bibinfo{pages}{1465--1468}.
\newblock


\bibitem[\protect\citeauthoryear{Waskom, Botvinnik, O'Kane, Hobson, Ostblom,
  Lukauskas, Gemperline, Augspurger, Halchenko, Cole, Warmenhoven, de~Ruiter,
  Pye, Hoyer, Vanderplas, Villalba, Kunter, Quintero, Bachant, Martin, Meyer,
  Miles, Ram, Brunner, Yarkoni, Williams, Evans, Fitzgerald, Brian, and
  Qalieh}{Waskom et~al\mbox{.}}{2018}]%
        {seaborn}
\bibfield{author}{\bibinfo{person}{Michael Waskom}, \bibinfo{person}{Olga
  Botvinnik}, \bibinfo{person}{Drew O'Kane}, \bibinfo{person}{Paul Hobson},
  \bibinfo{person}{Joel Ostblom}, \bibinfo{person}{Saulius Lukauskas},
  \bibinfo{person}{David~C Gemperline}, \bibinfo{person}{Tom Augspurger},
  \bibinfo{person}{Yaroslav Halchenko}, \bibinfo{person}{John~B. Cole},
  \bibinfo{person}{Jordi Warmenhoven}, \bibinfo{person}{Julian de Ruiter},
  \bibinfo{person}{Cameron Pye}, \bibinfo{person}{Stephan Hoyer},
  \bibinfo{person}{Jake Vanderplas}, \bibinfo{person}{Santi Villalba},
  \bibinfo{person}{Gero Kunter}, \bibinfo{person}{Eric Quintero},
  \bibinfo{person}{Pete Bachant}, \bibinfo{person}{Marcel Martin},
  \bibinfo{person}{Kyle Meyer}, \bibinfo{person}{Alistair Miles},
  \bibinfo{person}{Yoav Ram}, \bibinfo{person}{Thomas Brunner},
  \bibinfo{person}{Tal Yarkoni}, \bibinfo{person}{Mike~Lee Williams},
  \bibinfo{person}{Constantine Evans}, \bibinfo{person}{Clark Fitzgerald},
  \bibinfo{person}{Brian}, {and} \bibinfo{person}{Adel Qalieh}.}
  \bibinfo{year}{2018}\natexlab{}.
\newblock \bibinfo{title}{Seaborn: Statistical Data Visualization Using
  Matplotlib}.
\newblock
\newblock
\urldef\tempurl%
\url{https://doi.org/10.5281/zenodo.592845}
\showDOI{\tempurl}


\bibitem[\protect\citeauthoryear{Wilson, Bryan, Cranston, Kitzes, Nederbragt,
  and Teal}{Wilson et~al\mbox{.}}{2017}]%
        {Wilson2017}
\bibfield{author}{\bibinfo{person}{Greg Wilson}, \bibinfo{person}{Jennifer
  Bryan}, \bibinfo{person}{Karen Cranston}, \bibinfo{person}{Justin Kitzes},
  \bibinfo{person}{Lex Nederbragt}, {and} \bibinfo{person}{Tracy~K. Teal}.}
  \bibinfo{year}{2017}\natexlab{}.
\newblock \showarticletitle{{Good enough practices in scientific computing}}.
\newblock
  \bibinfo{howpublished}{\url{http://dx.plos.org/10.1371/journal.pcbi.1005510}}.
\newblock \bibinfo{journal}{\emph{PLOS Computational Biology}}
  \bibinfo{volume}{13}, \bibinfo{number}{6} (\bibinfo{date}{Jun}
  \bibinfo{year}{2017}), \bibinfo{pages}{e1005510}.
\newblock
\showISSN{1553-7358}
\urldef\tempurl%
\url{https://doi.org/10.1371/journal.pcbi.1005510}
\showDOI{\tempurl}


\end{thebibliography}
\clearpage

\section{Appendix}
\subsection{Dataset/corpus details}
\subsubsection{Keyword labels}
\label{a:keywords}
To capture the topical and disciplinary diversity of papers in our corpus, we assigned one or more keyword labels to each paper, intended to capture topical, domain, disciplinary, and methodological qualities about the study. A paper seeking to classify tweets for spam and phishing in Turkish might include the labels: spam detection; phishing detection; cybersecurity; non-English. A study seeking to classify whether users are tweeting in support or opposition of a protest might have the keywords: user profiling; political science; protests; stance detection; public opinion. As part of the annotation and labeling process, all five annotators gave each paper a short description of what was being classified or predicted. The project lead aggregated these independent descriptions and additionally examined the paper title, abstract, and text. The project lead --- who has extensive knowledge and experience of the various disciplines in the social computing space --- then conducted a two-stage thematic coding process. A first pass involved open (or free-form) coding for all papers, with the goal of creating a typology of keywords. The list of keywords were then refined and consolidated, and a second pass was conducted on all of the items to re-label them as appropriate. Papers could have multiple keywords. 

\begin{figure}
    \centering
    \includegraphics[width=0.5\textwidth]{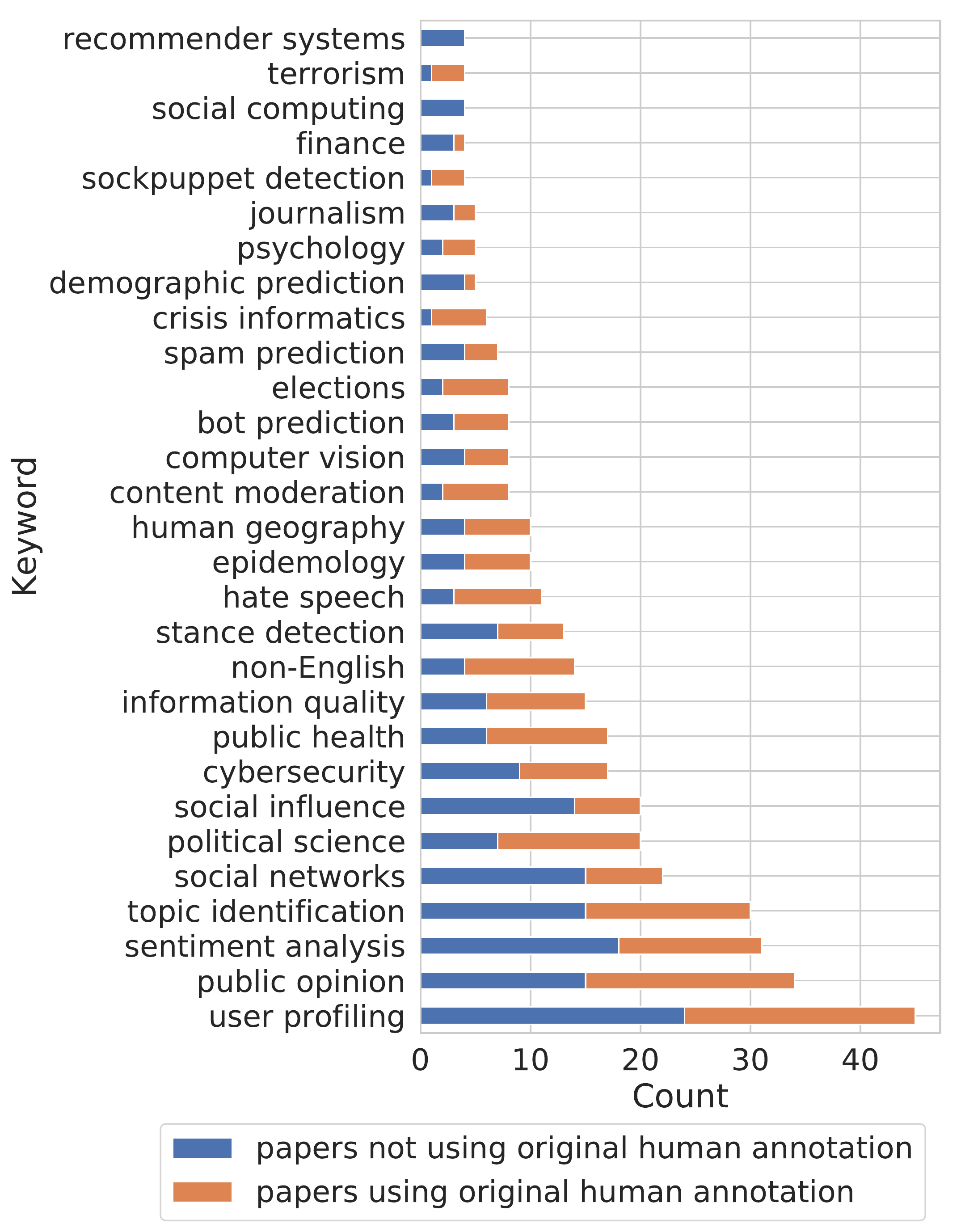}
    \caption{Plotting the distribution of papers by topical and disciplinary keywords, separated for papers using and not using original human annotation.}
    \label{fig:keyword-count}
\end{figure}

The distribution is plotted in Figure \ref{fig:keyword-count}, which is broken out by papers that were using original human annotation (e.g. a new labeled training dataset) versus either theoretical papers or papers exclusively re-using a public or external dataset (see section \ref{sec:original_human_annot}). This shows that the most common keywords were user profiling (a broader keyword that includes demographic prediction and classification of users into various categories), public opinion (a broader keyword that includes using Twitter to obtain beliefs or opinions, typically about political or cultural topics), and then two NLP methodologies of sentiment analysis and topic identification. The keyword "social networks" was used for any paper that either made substantive use of the network structure (e.g. follower graphs) as a feature, or tried to predict it. This figure also shows that our corpus also includes papers from a wide range of fields and sub-fields across disciplines, including a number of papers on cybersecurity (including bot/human detection, phishing detection, and spam detection), public health and epidemology, hate speech and content moderation, human geography, computer vision, political science, and crisis informatics. Papers using non-English languages were also represented in our corpus.

\subsubsection{Distribution of paper types in the corpus}
\label{a:papertypes}
\begin{table*}[h!]
\centering
\begin{tabular}{|p{4.5cm}|p{2cm}|l|l|p{2cm}|l|}
\hline
\textbf{} & Preprint never published & Postprint & Preprint & Non-ArXived (Scopus) & \textbf{Total} \\ \hline
Preprint never published & 57 & - & - & - & \textbf{57} \\ \hline
\begin{tabular}[c]{@{}l@{}}Refereed conference\\   proceedings\end{tabular} & - & 40 & 17 & 23 & \textbf{80} \\ \hline
Refereed journal article & - & 8 & 7 & 6 & \textbf{21} \\ \hline
Workshop paper & - & 2 & 3 & 0 & \textbf{5} \\ \hline
Dissertation & - & 1 & 0 & 0 & \textbf{1} \\ \hline
\textbf{Total} & \textbf{57} & \textbf{51} & \textbf{27} & \textbf{29} & 164 \\ \hline
\end{tabular}
\caption{Distribution of publication types in paper corpus.}
\label{tab:paper-types}
\end{table*}

For each of our 164 papers, we needed to determine various bibliometric factors. For papers in the ArXiv sample, the most important of these is whether the file uploaded to ArXiV is a version of a paper published in a more traditional venue, and if so, whether the ArXiV version is a pre-print submitted prior to peer-review (and has different content than the published version) or if it is a post-print that is identical in content to the published version. Many authors upload a paper to ArXiv when they submit it to a journal, others upload the accepted manuscript that has passed peer-review but has not been formatted and typeset by the publisher, and others upload the exact ``camera-ready'' version published by the publishers. ArXiV also lets authors update new versions; some will update each of these versions as they progress through the publishing process, others will only upload a final version, and some only upload the pre-review version and do not update the version in ArXiv to the published version.

To do this, the project lead first manually searched for the exact text of the title in Google Scholar, which consolidates multiple versions of papers with the same title. Papers that only had versions in ArXiv, ArXiv mirrors (such as adsabs), other e-print repositories like ResearchGate, personal websites, or institutional repositories were labeled as ``Preprint never published.'' For papers that also appeared in any kind of publication venue or publishing library (such as the ACM, IEEE, AAAI, or ACL digital libraries), the project lead recorded the publication venue and publisher, then downloaded the published version. In some workshops and smaller conferences, the ``publisher'' was a single website just for the event, which lacked ISSNs or DOIs. These were considered to be published as conference or workshop proceedings, if there was a public list of all the papers presented at the event with links to all of the papers. There was only one case in which there were two or more publications with the exact same title by the same authors, which involved a 2-page archived extended abstract for a poster in an earlier conference proceeding and a full paper in a later conference proceeding. For this case, we chose the full paper in the later venue.

The project lead then compared the version uploaded to ArXiv with the published version. As this was done after the labeling process, for papers where the author uploaded multiple versions to ArXiv, we took care to examine the version our labelers examined. If there were any differences in substantive content, the paper was labeled as ``Preprint of'' and then an appropriate description of the venue, such as ``refereed conference proceeding'' or ``refereed journal article.'' If there were no differences in the substantive content of the paper, the paper was labeled as ``Postprint of'' and then the venue description. Changes in reference style or ordering, page layout, typesetting, the size or color of figures, or moving the same text between footnotes and inline parentheticals were not considered to be substantive content changes. However, even a single character typo fix to the main body text, a single added or removed reference, or a change to a figure's caption constituted a substantive content change. Table \ref{tab:paper-types} shows the distribution of paper types. Because there was only one dissertation in the sample, which also was not using original human annotation, we excluded this category from the aggregate analyses by paper type shown in the results section.
\begin{table}[h!]
\begin{tabular}{@{}lp{2.75cm}p{2.75cm}}
\toprule
\textbf{Year} & \textbf{\# in ArXiv sample} & \textbf{\# in Scopus sample} \\ \midrule
2010 & 1 & 0 \\
2011 & 2 & 2 \\
2012 & 2 & 2 \\
2013 & 8 & 0 \\
2014 & 5 & 4 \\
2015 & 13 & 3 \\
2016 & 29 & 5 \\
2017 & 36 & 4 \\
2018 & 39 & 9 \\ \bottomrule
\end{tabular}
\caption{Count of publications per year}
\label{tab:year-counts}
\begin{tabular}{lllll}
\multicolumn{2}{l}{\textit{From ArXiv sample}} &  & \multicolumn{2}{l}{\textit{From Scopus sample}} \\ \midrule
\textbf{Publisher} & \textbf{Count} &  & \textbf{Publisher} & \textbf{Count} \\ \midrule
ArXiv-only & 58 &  & Springer & 7 \\
ACM & 20 &  & ACM & 5 \\
IEEE & 18 &  & Elsevier & 4 \\
Springer & 14 &  & SPC & 1 \\
ACL & 12 &  &  &  \\
Elsevier & 4 &  &  &  \\
AAAI & 3 &  &  &  \\
Sage & 1 &  &  &  \\
CEUR & 1 &  &  &  \\
PLoS & 1 &  &  &  \\
UIC & 1 &  &  &  \\
ISCRAM & 1 &  &  &  \\
JMIR & 1 &  &  &  \\ \bottomrule
\end{tabular}
\caption{Count of publishers from both samples}
\label{tab:publisher_counts}
\end{table}

\subsubsection{Distribution of publishers in corpus}
\label{a:pubtypes}
For each paper in the Scopus samples and each paper in the ArXiv corpus that was a pre-print or post-print of a published paper, we also collected information about the journal and publisher. There were 80 different journals, conference proceedings, or workshops represented, with the top venues being the proceedings of SocInfo with 6 papers and the proceedings of ASONAM (Advances in Social Network Analysis and Mining) with 4 papers. Six venues had 3 publications each, which were all conference proceedings: AAAI ICWSM, ELRA LREC, ACM CIKM, ACM WWW, and IEEE Big Data. The distribution of publishers is presented in table \ref{tab:publisher_counts}, which is broken out by papers in the ArXiv and Scopus corpus. The distribution of papers by years is shown in table \ref{tab:year-counts}.

\subsection{Methods and analysis details}

\subsubsection{Inter-annotator agreement}
\label{a:irr}
In the first round, 5 annotators examined each paper independently, then met to discuss papers with disagreement. Table \ref{table:irr} shows for each question, what percent of items were given the same label by all annotators (with number of annotators being recoded for the presence or absence of any information). Cases where no annotator answered the question because it was not relevant (e.g. crowdworker compensation for non-crowdworker projects) were not included in such a calculation, which would have increased such rates even more, but this would be somewhat disingenuous. 

We report percent complete agreement among all raters for each question; for each item, what percent were given the same rating by all raters? We believe this is a more appropriate and straightforward metric for our project. This is due to the fact that our data does not necessarily meet the particular assumptions of other widely used two statistical estimators for 3+ raters. Fleiss's kappa and Krippendorf's alpha are widely used because they take into account the possibilities that raters made decisions based on random chance. However, this requires assuming a uniform prior possibility of such a random distribution, which generally only applies if each possible response by raters is equally likely \cite{quarfoot_how_2016,oleinik_choice_2014}. This is the case in balanced datasets, but we observed widely skewed distributions. 

\begin{table*}[t]
\begin{tabular}{@{}lllll@{}}
\toprule
\textbf{Question} & \textbf{\% agreement, round 1} & \textbf{\% agreement, round 2} &   \\ \midrule
original classification task & 69.7\% & 93.9\% &   \\
labels from human annotation & 51.3\% & 82.9\% &   \\
used original human annotation & 72.0\% & 85.4\% &   \\
used external human annotation & 51.1\% & 63.4\% &   \\
original human annotation source & 44.3\% & 79.3\% &   \\
number of annotators & 38.2\% & 95.7\% &   \\
training for human annotators & 81.0\% & 84.8\% &   \\
formal instructions & 50.1\% & 82.9\% &   \\
prescreening for crowdwork platforms & 83.7\% & 89.0\% &   \\
multiple annotator overlap & 69.3\% & 81.7\% &   \\
reported inter-annotator agreement & 79.2\% & 83.5\% &   \\
reported crowdworker compensation & 94.9\% & 89.0\% &   \\
link to dataset available & 82.1\% & 86.0\% &   \\ \midrule
\textbf{Mean score} & 66.7\% & 84.4\% & \\
\textbf{Median score} & 69.5\% & 84.8\% & \\ \bottomrule

\end{tabular}
\caption{Inter-annotator reliability metrics (percent total agreement for all raters) for each question in both rounds of our process. Note that some items were added very late in round 1, including ``used external human annotation'', ``reported crowdworker compensation'' and ``link to dataset available.'' }
\label{table:irr}
\end{table*}

The rates of proportional agreement were not high enough in the first round for us to be confident, which is likely due to a variety of factors. First, in contrast to most of the papers we examined, our project involved annotators answering 13 different questions for each item, which adds significant complexity to the process. Second, machine learning publications are also some of the more difficult pieces of content to make determinations around, as the definitions and boundaries of various concepts are often relatively undefined and contested across the many academic disciplines. In particular, our lowest rate for the second round was in the external human annotation question, which was added between the first and second round, and appears to still have some ambiguity.

We observed substantial increases in agreement between round one and two, although this also is likely confounded by the fact that all five annotators reviewed every item in round one, but only two or three reviewed every item in round two. We should note that as our approach was a human annotation research project studying human annotation research projects, this has given us much empathy for how difficult such a task is. We also acknowledge that our project involves the same kind of ``black boxing'' we discussed in the literature review, in which a messy process of multiple rounds of human annotations is reduced to a gold standard. However, we do believe in being open about our process, and our data for both rounds of annotation and the final dataset will be available upon publication.

The overall question for any study involving structured human annotation is whether the entire annotation, integration, review, and reconciliation process ultimately results in high confidence for the final dataset. The standard approach of human annotation checked by inter-rater reliability treats individual humans as instruments that turn phenomena in the world into structured data. If there is a high degree of inter-rater reliability, then each individual human can generally be trusted to make the same determination. If this is the case, then either reconciliation can easily take place through a majority vote process involving no discussion, or if rates are quite high, then only a subset of items need to be reviewed multiple times. In contrast, what our first round of inter-rater reliability metrics told us was that we were not the same kinds of standardized instruments that turn the same inputs into the same outputs. This does not bode well if we were conducting a single-stage mechanical majority-rule reconciliation process, and certainly would be unwise if we only had a single individual annotate each paper. For such a reason, we did not rely on such easier processes of reconciliation and demanded all papers be annotated by multiple individuals and discussed in a group setting moderated by the lead research scientist.

Furthermore, because our approach was largely focused on identifying the presence of various kinds of information within long-form publications, this is a different kind of human judgment than is involved in common tasks using human annotators in social computing, such as social media content moderation, sentiment analysis, or image labeling. Typically, annotated items are much smaller and tend to be evaluated holistically, with disagreements arising from annotators who looked at the same information and made different determinations.

In contrast, we reflected that in our reconciliation process, most of the time when annotators disagreed, it was because some annotators had caught a piece of information in the paper that others had not seen. There was a common occurrence wherein one of the annotators would point out a particular paragraph, the other annotators who had initially disagreed would read it, and then remark that they had missed that part and would like to change their answer. That said, there were cases wherein annotators were reading the same sections of the paper and still arriving at different answers, which was often either 1) because the paper was giving ambiguous, incomplete, or implicit information, or 2) because there was a fundamental interpretation of the coding schema, which required updating the schema or the examples in it. For such reasons, we are relatively confident that if, after our two rounds of annotation and the reconciliation process, no individual member of our team has identified the presence of such information, then it is quite likely it is not present in the paper. 

\subsubsection{Changes to the coding schema}
\label{a:schema}
Unlike in some approaches to structured content analysis, the coding schema was open to revision if needed during this first round. Some difficult edge cases led to the refinement of the schema approximately half-way through this round of the labeling. The schema was developed on a web-based word processing platform, which also included examples of difficult edge cases, which were added as they were identified in team meetings. The document detailed each question, a formal definition or explanation of the question, the list of possible permitted labels, and various cases of examples that illustrated difficult or edge cases. 

The coding schema was modified only in cases where backward compatibility could be maintained with prior labeling work. This typically involved taking a question which had many granular possible labels and consolidating the possible labels into a smaller number of broader labels. For example, the question about whether instructions were given to human annotators originally involved specifying whether the instructions included a formal definition, examples, or both. This was revised to only specify ``instructions with formal definition or examples.'' Similarly, training for human annotators originally included a more granular list of possible training circumstances, plus ''no information'', ''other'', and ''unsure''. Because of the difficulty of gaining consensus on these different forms of training and the relatively few number of papers that gave any details whatsoever about annotator training (as well as no papers that explicitly stated no training had occurred), these were reduced to ``some training details'', ``no information'', and ''unsure'' (see Table \ref{table:schema_revisions}).

\begin{table}[h]
\begin{tabular}{@{}l|l@{}}
\textbf{Original coding schema} & \textbf{Revised coding schema} \\ \midrule
interactive training & some training details \\
professional training & some training details \\
prescreening with feedback & some training details \\
no training (explicitly stated) & other \\
other & other \\
no information & no information \\
unsure & unsure
\end{tabular}
\caption{Revisions in coding schema for question on annotator training}
\label{table:schema_revisions}
\end{table}

In addition, three questions were added halfway through the first round of the annotation process. First, a question was added about whether the paper used an external human-annotated dataset or not, which was added to clarify the question about whether original human annotation was used. This was added after a paper was discussed where an external human-annotated dataset was combined with an original human-annotated dataset. Two other questions were added about whether the paper contains a link to the training dataset and whether details about crowdworker compensation were included for projects using crowdworkers. These were both relatively straightforward questions, with relatively few incidences across our dataset. All papers had all questions answered in the second round.



\subsection{Software used}

All computational analysis and scripting was conducted in Python 3.7 \cite{python}, using the following libraries: Pandas dataframes \cite{pandas} for data parsing and transformation; SciPy \cite{scipy} and NumPy \cite{numpy} for quantitative computations; and Matplotlib \cite{Matplotlib} and Seaborn \cite{seaborn} for visualization. Analysis was conducted in Jupyter Notebooks \cite{jupyter} using the IPython \cite{ipython} kernels. Datasets and Jupyter Notebooks for data collection and analysis will be made available upon publication, which are made to run on Binder \cite{binder}.

\subsection{Coding schema, examples, and instructions}
\label{a:instructions}
A final version of our coding schema and instructions is below:

\setlength{\parindent}{0pt}
\addtolength{\parskip}{6pt}
\setlength{\intextsep}{6pt}

\textbf{1. Original classification task:} Is the paper presenting its own original classifier that is trying to predict something? ``Original'' means a new classifier they made based on new or old data, not anything about the novelty or innovation in the problem area.

Machine learning involves any process that does not have explicit or formal rules, where performance increases with more data. Classification involves predicting cases on a defined set of categories. Prediction is required, but not enough. Linear regressions might be included if the regression is used to make a classification, but making predictions for a linear variable is not. Predicting income or age brackets is classification, predicting raw income or age is not.

\begin{itemize}
 
\item Example: analyzing statistics about the kinds of words people use on
social media is not a classification task at all.

\item Example: predicting location is a classification task if it is from
work, school, home, or other, but not if it is an infinite/undefined
number of locations.

\item Example: This paper (https://ieeexplore.ieee.org/document/7937783) was
framed as not an original classification task (more algorithm
performance), but they did create an original classifier. This can also
be an ``unsure'' -- which is 100\% OK to answer.

\item Example: Literature review papers that include classification papers
aren't in this, if they didn't actually build a classifier.

\item Example: if there is a supervised classification task that is part of a
broader process, this counts, focus on that.

\end{itemize}

If no, skip the following questions.

\textbf{2. Classification outcome:} What is the general type of problem
or outcome that the classifier is trying to predict? Keep it short if
possible. For example: sentiment, gender, human/bot, hate speech,
political affiliation.

\textbf{3. Labels from human annotation:} Is the classifier at least in
part trained on labeled data that humans made for the purpose of the
classification problem? This includes re-using existing data from human
judgments, if it was for the same purpose as the classifier. This does
not include clever re-using of metadata.

Do a quick CTRL-F for ``manual'' and ``annot'' if you don't see
anything, just to be sure.

If not, skip the following questions about human annotation.
\begin{itemize}
  
\item Example: ISideWith paper on political stances was labels from human
annotation, just not original. They took the labels from elsewhere and
filled in the gaps (more on that in next Q).

\item Example: Buying followers and seeing who follows (1411.4299.pdf) is not
human annotation.

\item Example: Generating (smart) simulated datasets from metadata is not
human annotation.

\item Example: 1612.08207.pdf is not annotation when looking up political
affiliation of politicians from an external database, even though it is
manual work. No judgment is involved.

\item Example: 1709.01895.pdf is labels from human annotation, even though it
is semi-automated. They identified hashtags that they believe
universally correspond to certain political stances. There is a form of
human judgment here, although in that paper, they don't define or
explain it.

\item Example: Evaluation using human annotation is not annotation for ML, if
the annotation wasn't used to make the classifier. (1710.07394.pdf)

\item Example: If they are using human annotation just to have confidence that
a machine-annotated dataset is as good as a human annotated one, but the
human annotated dataset isn't actually used to train the classifier, it
is *not* using human annotation for ML. (1605.05195.pdf)

\end{itemize}

\textbf{4. Used original human annotation:} Did the project involve
creating new human-labeled data, or was it exclusively re-using an
existing dataset?

\begin{itemize}
\item
  \begin{quote}
  Yes
  \end{quote}
\item
  \begin{quote}
  No
  \end{quote}
\item
  \begin{quote}
  Unsure
  \end{quote}
\end{itemize}

Papers may have a mix of new and old human labeled data, or new human
labeled data and non-human labeled data. If there is any new human
annotation, say yes.

New human annotation must be systematic, not filling in the gaps of
another dataset. Example: ISideWith paper on political stances is *not*
original human annotation, even though they did some manual original
research to fill the gap.

If the methods section is too vague to not tell, then leave as unsure
(example: 1801.06294.pdf)

\textbf{4.5. Used external human annotation data:} Did the project use
an already existing dataset from human labeled data?

\begin{itemize}
\item
  \begin{quote}
  Yes
  \end{quote}
\item
  \begin{quote}
  No
  \end{quote}
\item
  \begin{quote}
  Unsure
  \end{quote}
\end{itemize}

If they are using external human annotated data, skip the remaining
questions:

\textbf{5. Original human annotation source:} Who were the human
annotators? Drop-down options are:

\begin{itemize}
\item
  \begin{quote}
  Amazon Mechanical Turk (AMT, Turkers)
  \end{quote}
\item
  \begin{quote}
  Any other crowdworking platform (Crowdflower / Figure8)
  \end{quote}
\item
  \begin{quote}
  The paper's authors
  \end{quote}
\item
  \begin{quote}
  Academic experts / professionals in the area
  \end{quote}
\item
  \begin{quote}
  No information in the paper
  \end{quote}
\item
  \begin{quote}
  Other
  \end{quote}
\item
  \begin{quote}
  Unsure
  \end{quote}
\end{itemize}

For academic experts or professionals in the area, this is independent
from the kinds of specific training they received for the task at hand.
Think of ``the area'' broadly, so if it is something about healthcare
and nurses were recruited, that would be professionals in the area, even
if they don't say anything about the nurses having specific training in
the annotation task at hand. If it doesn't easily fit into these or uses
multiple sources, add them in the next column.

\begin{itemize}
\item Example: ``We develop a mechanism to help three volunteers analyze each
collected user manually'' -\/- put other, if that is all they say

\item Example: If it just says ``we annotated\ldots{}'' then assume it is only
the paper's authors unless otherwise stated.
\end{itemize}

\textbf{6. Number of human annotators:}

Put the number if stated, if not, leave blank.

\textbf{7. Training for human annotators:} Did the annotators receive
interactive training for this specific annotation task / research
project? Training involves some kind of interactive feedback. Simply
being given formal instructions or guidelines is not training. Prior
professional expertise is not training. Options include:

\begin{itemize}
\item
  \begin{quote}
  Some kind of training is mentioned
  \end{quote}
\item
  \begin{quote}
  No information in the paper
  \end{quote}
\item
  \begin{quote}
  Unsure
  \end{quote}
\end{itemize}

Example: It is not considered training if there was prescreening, unless
they were told what they got right and wrong or other debriefing. Not
training if they just gave people with high accuracy more work.

Example: This paper had a minimum acceptable statement for some training
information, with only these lines: ``The labeling was done by four
volunteers, who were carefully instructed on the definitions in Section
3. The volunteers agree on more than 90\% of the labels, and any
labeling differences in the remaining accounts are resolved by
consensus.''

\textbf{8. Formal instructions/guidelines:} What documents were the
annotators given to help them? This document you are in right now is an
example of formal instructions with definitions and examples.

\begin{itemize}
\item
  \begin{quote}
  No instructions beyond question text
  \end{quote}
\item
  \begin{quote}
  Instructions include formal definition or examples
  \end{quote}
\item
  \begin{quote}
  No information in paper (or not enough to decide)
  \end{quote}
\item
  \begin{quote}
  Unsure
  \end{quote}
\end{itemize}

Example of a paper showing examples: ``we asked crowdsourcing workers to
assign the `relevant' label if the tweet conveys/reports information
useful for crisis response such as a report of injured or dead people,
some kind of infrastructure damage, urgent needs of affected people,
donations requests or offers, otherwise assign the `non-relevant'
label''

\textbf{9. Prescreening for crowdwork platforms }

Leave blank if this is not applicable.

\begin{itemize}
\item
  \begin{quote}
  No prescreening (must state this)
  \end{quote}
\item
  \begin{quote}
  Previous platform performance qualification (e.g. AMT Master)
  \end{quote}
\item
  \begin{quote}
  Generic skills-based qualification (e.g. AMT Premium)
  \end{quote}
\item
  \begin{quote}
  Location qualification
  \end{quote}
\item
  \begin{quote}
  Project-specific prescreening: researchers had known ground truth and
  only invited
  \end{quote}
\item
  \begin{quote}
  No information
  \end{quote}
\item
  \begin{quote}
  Unsure
  \end{quote}
\end{itemize}

\textbf{10. Multiple annotator overlap:} Did the annotators label at
least some of the same items?

\begin{itemize}
\item
  \begin{quote}
  Yes, for all items
  \end{quote}
\item
  \begin{quote}
  Yes, for some items
  \end{quote}
\item
  \begin{quote}
  No
  \end{quote}
\item
  \begin{quote}
  Unsure
  \end{quote}
\item
  \begin{quote}
  No information
  \end{quote}
\end{itemize}

If it says there was overlap but not info to say all or some, put
unsure.

\textbf{11. Reported inter-annotator agreement:} Leave blank if there
was no overlap. Is a metric of inter-annotator agreement or intercoder
reliability reported? It may be called Krippendorf's alpha, Cohen's
kappa, F1 score, or other things.

\begin{itemize}
\item
  \begin{quote}
  Yes
  \end{quote}
\item
  \begin{quote}
  No
  \end{quote}
\item
  \begin{quote}
  Unsure
  \end{quote}
\end{itemize}

\textbf{12. Reported crowdworker compensation:} If using crowdworkers to
annotate, did they say how much the annotators were paid for their work?
Leave blank if crowdworkers were not used.

\begin{itemize}
\item
  \begin{quote}
  Yes
  \end{quote}
\item
  \begin{quote}
  No
  \end{quote}
\item
  \begin{quote}
  Unsure
  \end{quote}
\end{itemize}

\textbf{13. Link to dataset available:} Is there a link in the paper to
the dataset they used?

\begin{itemize}
\item
  \begin{quote}
  Yes
  \end{quote}
\item
  \begin{quote}
  No
  \end{quote}
\item
  \begin{quote}
  Unsure
  \end{quote}
\end{itemize}

\end{document}